\documentclass[intlimits,twoside,a4paper]{article}

\usepackage[eqsecnum]{cmpj3}
\usepackage[utf8]{inputenc}

\usepackage{bm}

\issue{2026}{29}{3}{33502}
\doinumber{10.5488/CMP.29.33502}

\title[Molecular dynamics simulations of DMSO--MeOH liquid mixtures]
{Molecular dynamics simulations of DMSO--MeOH liquid mixtures. Effects of force fields on mixing properties 
}

\author[M. Cruz-S\'{a}nchez,   O. Pizio]
{M. Cruz-S\'{a}nchez\orcid{0009-0005-0407-6496}\refaddr{label1},
O. Pizio\orcid{0000-0001-8333-4652}\refaddr{label2}
\thanks{Corresponding author: \email{oapizio@gmail.com}.}}
\addresses{
\addr{label1}
Departamento de Qu\'{i}mica, Universidad Aut\'{o}noma Metropolitana-Iztapalapa,
Av. San Rafael Atlixco 186, Col. Vicentina, 09340 CDMX, M\'{e}xico
\addr{label2}Instituto de Qu\'{i}mica, Universidad Nacional Aut\'{o}noma de M\'{e}xico,
Circuito Exterior, 04510 Cd. Mx., M\'{e}xico
}


\date{Received 29 June 2026; revised 28 July 2026; accepted 31 July 2026; published 28 September 2026}

\begin{document}

\maketitle

\begin{abstract}
	
We explore composition dependence of the principal properties of liquid dimethylsulfoxide (DMSO)-methanol (MeOH) liquid  mixtures by using molecular dynamics computer simulations. A set of non-polarizable semi-flexible models for the DMSO molecule combined with methanol models is investigated. Composition trends of density, excess mixing volume and excess mixing enthalpy are evaluated. Besides, we study the composition dependence of self-diffusion of species and shear viscosity ofthe static dielectric constant and the surface tension.Certain aspects of the microscopic structure  are analyzedin terms of radial distribution functions and of the average number of hydrogen-bonded molecules. The quality of several combinations of the models is illustrated and critically evaluated by comparisons with experimental data.

\printkeywords
%
%

\end{abstract}

\section{Introduction}

There have been  many experimental and computer simulation reports concerning measurements 
and interpretation of properties of binary mixtures involving dimethylsulfoxide
as one of the components.
This research was motivated by the practical importance of pure DMSO as a solvent
in chemistry and chemical engineering~\cite{jacobs,martin1,martin2}.
Moreover,  liquid mixtures of dimethylsulfoxide (DMSO) and water
are used in several  biological applications, the cryoprotection being just one of the
examples~\cite{mcgann,murthy,rabin,markarian}. Ionic solutes in
water--DMSO mixed solvents have been studied for diverse practical 
objectives as well~\cite{yan,bergman,xie1}.
As discussed in~\cite{wiewior} for example, the experimental knowledge and understanding
of the microscopic structure and dynamic properties of these systems 
follows mainly  from the application of calorimetry, neutron scattering,
nuclear magnetic resonance, dielectric relaxation and vibrational spectroscopy methods.
Hence, a set of experimental data for this class of mixtures is quite comprehensive. 

Less attention has been paid to mixtures of DMSO (and its analogues, like diethylsulfoxide) 
with non-aqueous solvents. These systems, however, are of both practical importance 
and academic interest as documented 
in several publications, see e.g.,~\cite{alfred,zeidler,virk,chaban,polyakov,rao,ghosh}. 
Finally, multi-component mixtures that involve
salts and/or room temperature ionic liquids besides DMSO and/or other species (e.g., water), 
were investigated recently~\cite{prausnitz,santos,wen,long,laria,ludwig}.

In order to interpret the experimental results in detail
and to get ampler insights with predictive power, one is
forced to resort to computer simulation methodology.
Usual strategy of computer simulations  is to choose a model of each species,
i.e. the intra-molecular structure at a certain level of sophistication, 
and assume the inter-molecular interactions. Commonly, the combination rules
are used for the cross interaction between species.

Most frequently used and quite well tested non-polarizable  models for monohydric alcohols 
are the ones at a united atom and all-atom levels, 
see e.g.,~\cite{jorgensen,trappe,salgado,melgarejo,jorgensen2}.
Similar united atom level models for DMSO species were proposed in the original publications
(cited below) and reviewed in detail in \cite{chalaris,idrissi}.
A more sophisticated force fields for DMSO
were investigated as well~\cite{strader,bachmann}. Namely, a flexible all-atom
model of DMSO was developed in \cite{strader} and tested 
in a few aspects for DMSO-water (TIP3P) mixtures.
On the other hand, Bachmann and van Gunsteren,~\cite{bachmann}, proposed
computationally demanding,  polarizable model for DMSO and combined it with polarizable water model.
Extensive tests for  DMSO--water systems were performed and suggested that simpler, non-polarizable
models with successful parametrization preserve competitive level.

Adequacy of the computer simulation predictions for a given model should be tested 
by comparison with experimental data. In general terms, computer simulations
permit to interpret observables at microscopic level with molecular level details.
Besides, a wider set of properties, in comparison with experimental results, can be obtained
with a reasonable degree of confidence.

In the design of alcohol models at a united atom level, e.g., the UAMI-EW 
set of models from \cite{melgarejo},
the target properties for fitting of the force field are the density, the surface tension
and the dielectric constant. Throughout this report we use abbreviation UAMI rather than UAMI-EW
for the sake of brevity.
This kind of parametrization was quite successful for a single-component
monohydric alcohols on temperature and pressure,  and for their mixtures with water
at a room temperature and ambient pressure~\cite{melgarejo,bako1,mine1}. However, our recent
study elucidated deficiencies of this type of force field upon its application
for mixtures of different alcohols~\cite{mine2}. Similar problem has been encountered 
while using TraPPE force field for mixtures of primary and secondary alcohols~\cite{siepmann}.

Concerning the DMSO modelling at a united atom level, one should note that most common targets
include solely the liquid density and the vaporization enthalpy, see e.g.,~\cite{geerke}.
Certain adjustments of parameters to yield various properties were used in some cases
for the sake of better performance as discussed in \cite{chalaris} for example.
Still, the principal attention of many studies was to apply the pure DMSO united atom models
and explore the properties of DMSO--water mixtures.

In spite of availability of experimental data for various properties of  DMSO--alcohol mixtures, 
we have found the only previous investigation of binary DMSO--methanol system by using 
molecular dynamics simulation in \cite{vechi}. 
These authors used the OPLS united atom type models for both species, MeOH and DMSO,
respectively. Principal findings of this work have been restricted to the exploration of
composition changes of microscopic structure, self-diffusion coefficients and molecular 
reorientation times without comprehensive comparisons with experimental predictions. 

Having this in mind, the principal objective of the present work is to investigate, 
as comprehensively as possible,
the mixing properties of methanol and DMSO by using a wide set of united-atom models for both species 
and molecular dynamics methodology. We would like to critically evaluate the results of simulations
versus experimental data in order to proceed to a more sophisticated modelling in future work and to attempt an
extension for a more complex alcohol species.

\section{Models and simulation details}

In this work we restrict our attention to a set of DMSO united atom type, 
non-polarizable  models with four sites, O, S, CH$_3$ (methyl group consisting of carbon
and hydrogen atoms is replaced by a single CH$_3$ interaction site
located on the methyl carbon atom).  The CH$_3$ group is abbreviated as C in what follows.
Within this modelling,
the interaction  potential between all atoms and/or groups is assumed
as a sum of Lennard-Jones (LJ) and Coulomb terms.
Similarly, the united atom type modelling is considered for methanol species. For this
subsystem we restrict our attention to the UAMI model and TraPPE force field.
Both of them were previously tested on several occasions to reproduce properties
of monohydric alcohols--water mixtures.

A comprehensive description of the parameters of models of this type
is given in tables~\ref{tab1}--\ref{tab3} below. The nomenclature of models is taken from the original
publications indicated in the first column of table~\ref{tab1}.

\begin{table}[h!]
  \small
    \caption{Parameters of Lennard-Jones site-site interaction for DMSO models.
	 The energies are given in kJ/$\text{mol}$ whereas the diameters are in \AA. }
    \vspace{0.1cm}
     \begin{center}
      \begin{tabular}{l c c c c c c c c c }
      \hline
       Model  &    $\varepsilon_{\text{OO}}$ & $\sigma_{\text{OO}}$ &    $\varepsilon_{\text{SS}}$ & $\sigma_{\text{SS}}$ &
                 $\varepsilon_{\text{CC}}$ & $\sigma_{\text{CC}}$  \\
       \hline
	      P2 \cite{luzar}    &  0.2992 & 2.80 & 0.9974 &  3.40  & 1.230 & 3.80   \\
	      OPLS \cite{zheng}  &  1.171  & 2.93 & 1.652  &  3.56  & 0.669 & 3.81   \\
	      VB \cite{vaisman}  &  0.276  & 2.94 & 0.842  &  3.56  & 0.669 & 3.60   \\	      
	      VLL \cite{vishnyakov}  &  0.591  & 2.92 & 1.398  &  3.66  & 0.957 & 3.76   \\	      
	      GOVG \cite{geerke}  &  1.7154  & 2.63 & 1.2972  &  3.56  & 0.86718 & 3.748   \\
	      Bordat et al. \cite{bordat} &  1.7154 & 2.63 & 1.297   &  3.56  & 0.9414  & 3.739   \\
       \hline
  \end{tabular}
  \end{center}
\label{tab1}
\end{table}
Similar information and nomenclature for DMSO models, the reader can find in
table~1 of \cite{chalaris}, in table~1  of \cite{idrissi}, and
in \cite{geerke,vishnyakov}.
In our table, some misprints from table 1 of~\cite{idrissi} were corrected. The most important misprint 
in the original paper \cite{vishnyakov} was the value for $\varepsilon_{\text{CC}}$, 
which was discovered first by Geerke et al.~\cite{geerke}. 
In addition, we take the original value for $\sigma_{\text{CC}}$ from \cite{vishnyakov}, in contrast to \cite{idrissi}.

\begin{table}[h!]
  \small
	  \caption{Charges of sites for the united atom DMSO models (in units of elementary charge).
          }
     \begin{center}
      \begin{tabular}{l c c c c }
      \hline
       Model  &    $q_{\text{O}}$  & $q_{\text{S}}$  &   $q_{\text{C}}$ & 
                  \\
       \hline
      P2, OPLS, VB, VLL   &  $-0.459$  & 0.139  & 0.16  \\
      GOVG      & $-0.44753$   & 0.12753 & 0.16   \\
      Bordat     & $-0.43674$   & 0.11674 & 0.16   \\
       \hline
  \end{tabular}
  \end{center}
\label{tab2}
\end{table}

Finally, the parameters describing intramolecular structure of DMSO molecule are:
$l_{\text{OS}} = 1.53$ \AA~and $l_{\text{SC}} = 1.80$ \AA~(P2, OPLS, VLL models); 
$l_{\text{OS}} = 1.53$ \AA~and $l_{\text{SC}} = 1.95$ \AA~(Bordat et al. model);
$l_{\text{OS}} = 1.53$ \AA~and $l_{\text{SC}} = 1.93799$ \AA~(GOVG model) and
$l_{\text{OS}} = 1.496$ \AA~and $l_{\text{SC}} = 1.80$ \AA~(VB model), respectively.
The angles are as follows O-S-C = 106.75$^{\circ}$, C-S-C = 97.4$^{\circ}$ 
(for P2, OPLS, VLL, GOVG  and Bordat et al.);  107.2$^{\circ}$, 
and 99.2$^{\circ}$ (for VB model),
respectively.
In addition, the improper dihedral S-C-O-C is included into the DMSO models according
to \cite{oostenbrink}. This issue was discussed in our previous studies of
water--DMSO mixtures~\cite{gujt,aguilar}.

  \begin{table}[h!]
  \small
    \caption{Parameters of Lennard-Jones site-site interactions for methanol models.
          The C-O-H angle is 108.5$^{\circ}$; $l_{\text{OH}} = 0.945$ \AA, $l_{\text{CO}} = 1.43$ \AA.
          }
    \vspace{0.1cm}
     \begin{center}
      \begin{tabular}{l c c c c c c c c  }
      \hline
       Model  &    $\varepsilon_{\text{OO}}$ & $\sigma_{\text{OO}}$ &    $\varepsilon_{\text{CC}}$ & $\sigma_{\text{CC}}$ &
                 $q_{\text{O}}$ & $q_{\text{C}}$ & $q_{\text{H}}$  \\
       \hline
      TraPPE    &  0.773245  & 3.02   & 0.814817  &  3.75    & $-0.700$   & 0.265   & 0.435  \\
      UAMI      &  0.68208   & 3.1808 & 0.71877   &  3.6783  & $-0.77147$ & 0.29252 & 0.47895 \\
       \hline
  \end{tabular}
  \end{center}
\label{tab3}
\end{table}

Molecular dynamics computer simulations of DMSO--methanol mixtures have been performed in the
isothermal-isobaric (NPT) ensemble at atmospheric  pressure 1~bar and at temperature 298.15~K.
We used GROMACS package~\cite{gromacs} version 5.1.2.
The simulation box in each run was cubic, the total number of molecules is
fixed at 3000. Composition of the mixture is described by the mole fraction of DMSO
molecules, $X_\text{D}=N_{\text{DMSO}}/(N_{\text{DMSO}}+N_{\text{MeOH}})$.
As common, periodic boundary conditions were used.
Temperature and pressure control was provided by the V-rescale thermostat and Parrinello-Rahman
barostat with $\tau_T$ = 0.5 ps and $\tau_P$ = 2.0 ps, the timestep was 0.002 ps.
The value of $5.25\cdot10^{-5}$ bar$^{-1}$ was used for the compressibility of mixtures.
Lorentz-Berthelot combination rules (CR2 in Gromacs nomenclature) 
were used to determine cross parameters for
the relevant potential well depths and diameters for all models.
The models for MeOH, UAMI and TraPPE, imply application of CR2 rules. Moreover,
the P2, VLL, GOVG and Bordat et al. models for DMSO were studied by using the CR2 rules
as well. Solely, the OPLS modelling requires the application of CR3 rules.
In summary, in the present study we restrict only to the CR2 procedure for all models, in close 
similarity to the study of the entire set of DMSO--water models (with CR2 
rules) in \cite{idrissi}. 

The non-bonded interactions were cut-off at 1.1 nm, whereas the long-range electrostatic interactions
were handled by the particle mesh Ewald method implemented in the GROMACS software package  (fourth
order, Fourier spacing equal to 0.12) with the precision $10^{-5}$.
The van der Waals correction terms to the energy and pressure were applied.
In order to maintain the geometry of molecules, the LINCS
algorithm was used.

After preprocessing and equilibration, several consecutive simulation runs
with
the starting configuration being the last configuration from the previous
run, were performed to obtain trajectories for the data analysis.
The results for each property  were obtained by averaging over pieces 
(20--30~ns) of the entire trajectories not less than 150 ns.
Additional details concerning calculations of the surface tension and shear
viscosity are given in the respective sections below.

\section{Results and discussion}

Nikam et al. \cite{nikam} reported experimental results for
the composition dependence of density and viscosity of 
DMSO-alkanols (methanol, ethanol, propanol, etc.) mixtures at 298.15~K and 303.15~K.
Besides, these authors 
provided  the excess mixing volume and the excess viscosity (in figures~\ref{fig2} and \ref{fig3} of~\cite{nikam}).
Previous simulation work,  \cite{vechi}, does not report these properties.

\subsection{Density and molar volume of DMSO--MeOH mixtures on composition}

\begin{figure}[h]
\begin{center}
\includegraphics[width=5.7cm,clip]{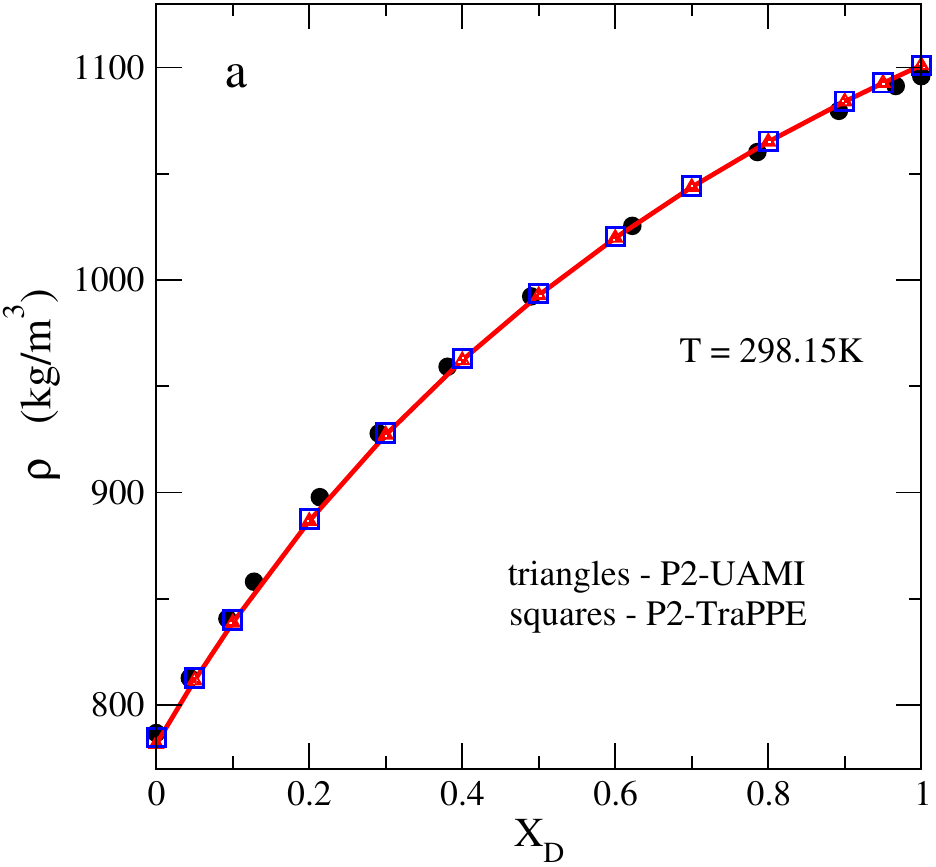}
\includegraphics[width=5.5cm,clip]{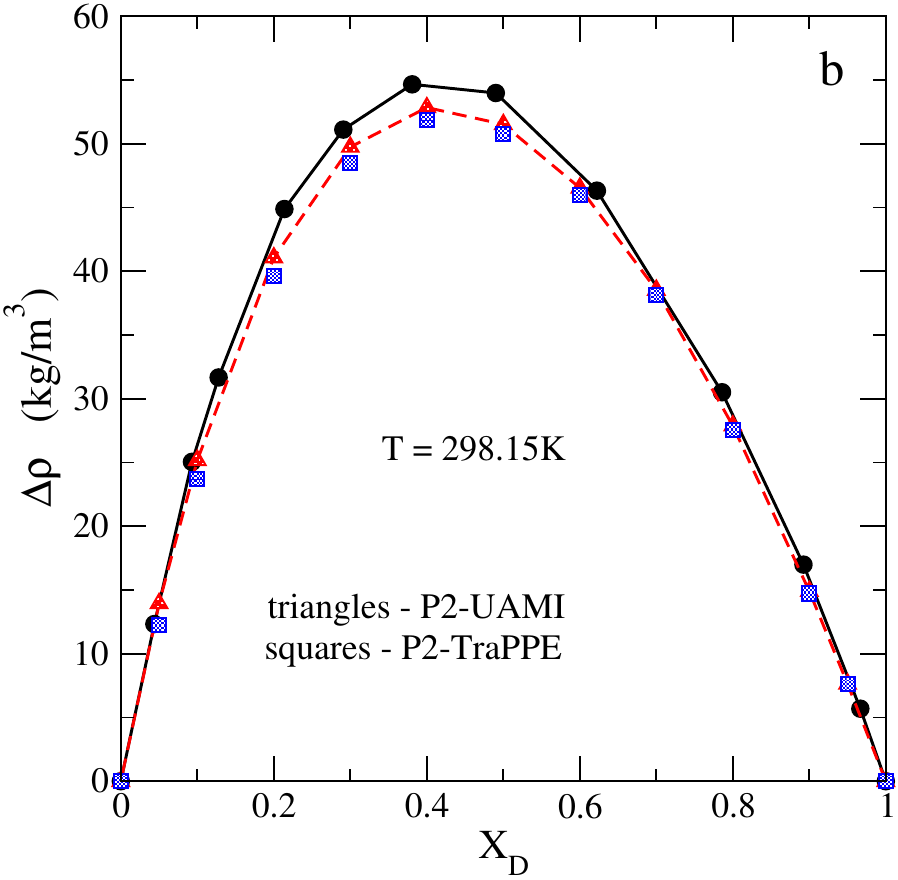}
\end{center}
\caption{(Colour online) Panels a and b:  
The dependence of DMSO--MeOH density (panel a) and 
of the excess density (panel b) 
on mole fraction of DMSO, $X_{\text D}$.
The simulation results are for P2--UAMI (red) and for P2--TraPPE (blue) united atom models.
The experimental data (black circles) are from \cite{nikam}.
The nomenclature of symbols is given in both panels of the figure.
All the results in the figures throughout this work refer to $T = 298.15$~K and $P = 1$~bar,
unless specified.
}
\protect
\label{fig1}
\end{figure}

We use experimental data of Nikam et al. at a room temperature
$T= 298.15$~K, and atmospheric pressure~\cite{nikam}.
In the first figure, figure~\ref{fig1}, we show the dependence of density on composition 
for a single DMSO model, P2 from \cite{luzar}, combined with either UAMI \cite{melgarejo}
or TraPPE \cite{trappe}  MeOH models. 
The P2 model for DMSO was chosen quite arbitrarily, mainly to keep track of
our previous studies of DMSO--water mixtures, \cite{gujt,aguilar}.

Previously, it was shown that the principal improvement achieved by the
UAMI methanol model, in comparison to the TraPPE one, for pure methanol, is in a much better
dielectric constant of the former, whereas other properties, e.g., density are
of the same quality.
Similar conclusion, concerning $\rho (X_{D})$ for DMSO--MeOH mixtures, is illustrated
in figure~\ref{fig1}a. The simulation data accurately reproduce experimental behavior of density on
composition. 
Moreover, both methanol models, if combined with the P2 DMSO force field,
yield correct and quite accurate description of the deviation of density of the mixture
from ideality ($\Delta\rho = \rho - X_{\text D}\rho_D -(1-X_{\text D})\rho_M$) in the entire composition range, figure~\ref{fig1}b.

In the following figure, figure~\ref{fig2}, we consider a single methanol model, UAMI, 
and explore the composition dependence of density for various DMSO models described in table~\ref{tab1}.
It can be seen that the VLL--UAMI model yields most accurate predictions.
On the other hand, the VB--UAMI model is the least accurate, it leads to a
qualitatively correct predictions but deviates from experimental points in the
interval of intermediate composition.

\begin{figure}[h!]
\begin{center}
\includegraphics[width=6.2cm,clip]{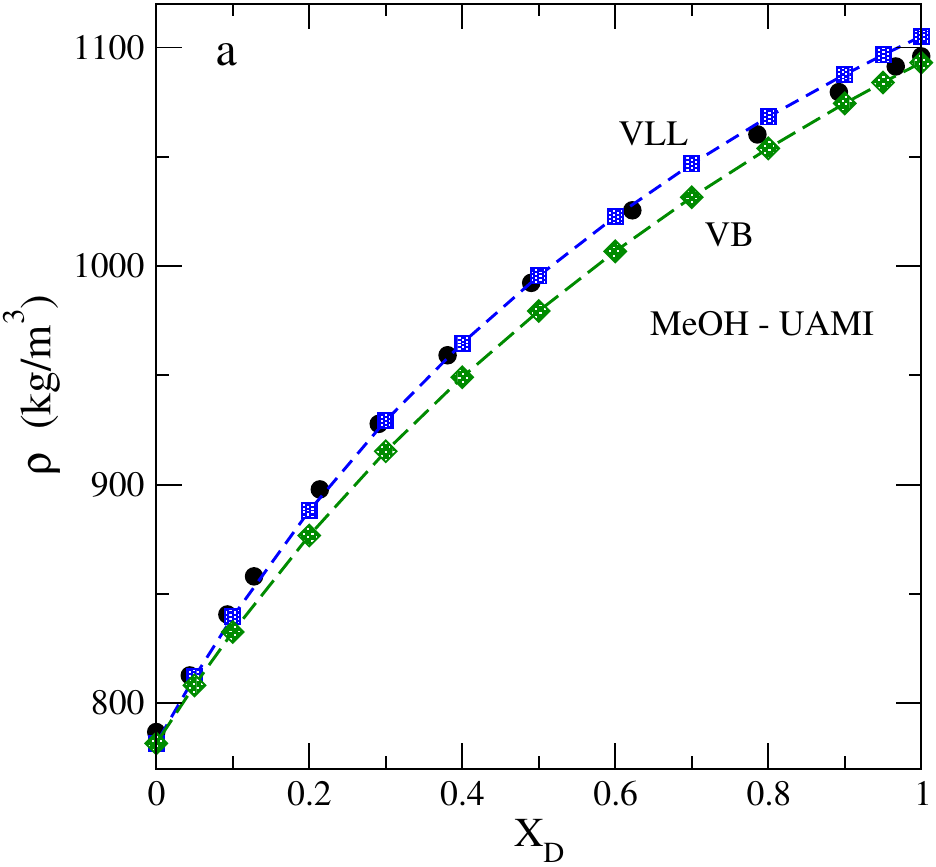}
\includegraphics[width=6cm,clip]{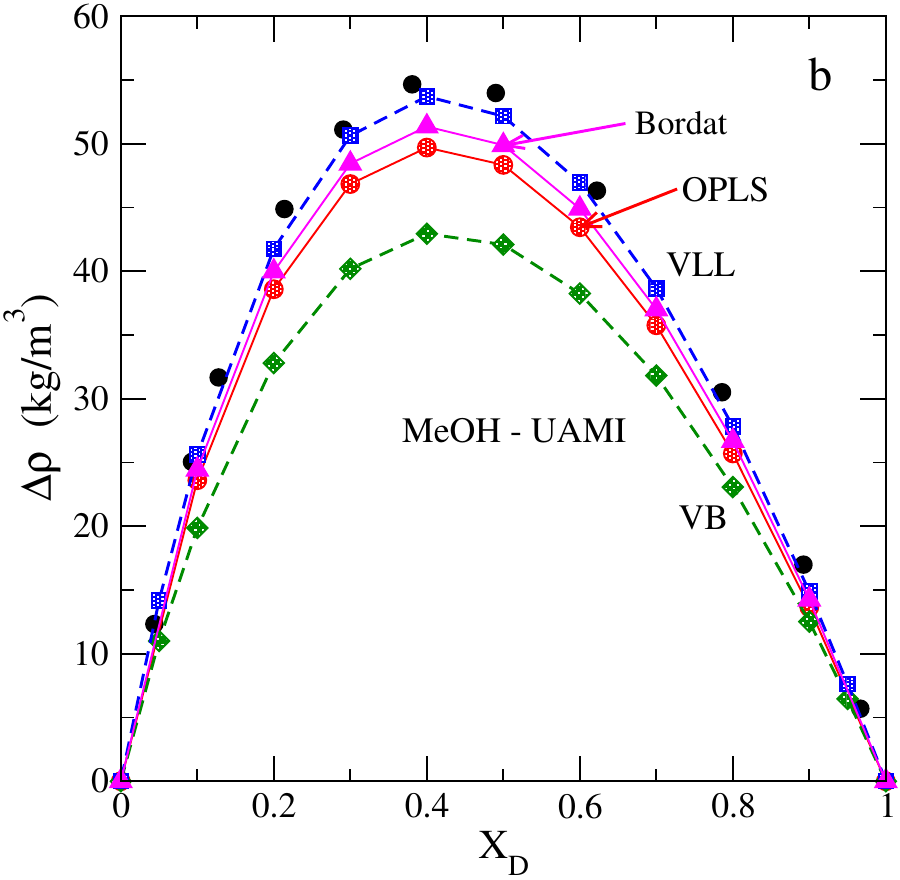}
\end{center}
\caption{(Colour online) Panels a and b:
The dependence of DMSO--MeOH density (panel a) and
of the excess density (panel b) 
on DMSO mole fraction, $X_{\text D}$. 
The experimental data (black circles) are from~\cite{nikam}. The nomenclature of colors and symbols is
given in both panels of the figure.
}
\protect
\label{fig2}
\end{figure}

Frequently, for chemical engineering purposes, the experimental data are reported 
in terms of the excess mixing volume, $\Delta V_{\text{mix}}$. 
It is defined by the same expression as $\Delta \rho$ above, just for molar volumes
rather than the densities, see e.g., equation~(2) of \cite{torres}. 

\begin{figure}[h!]
\begin{center}
\includegraphics[width=6cm,clip]{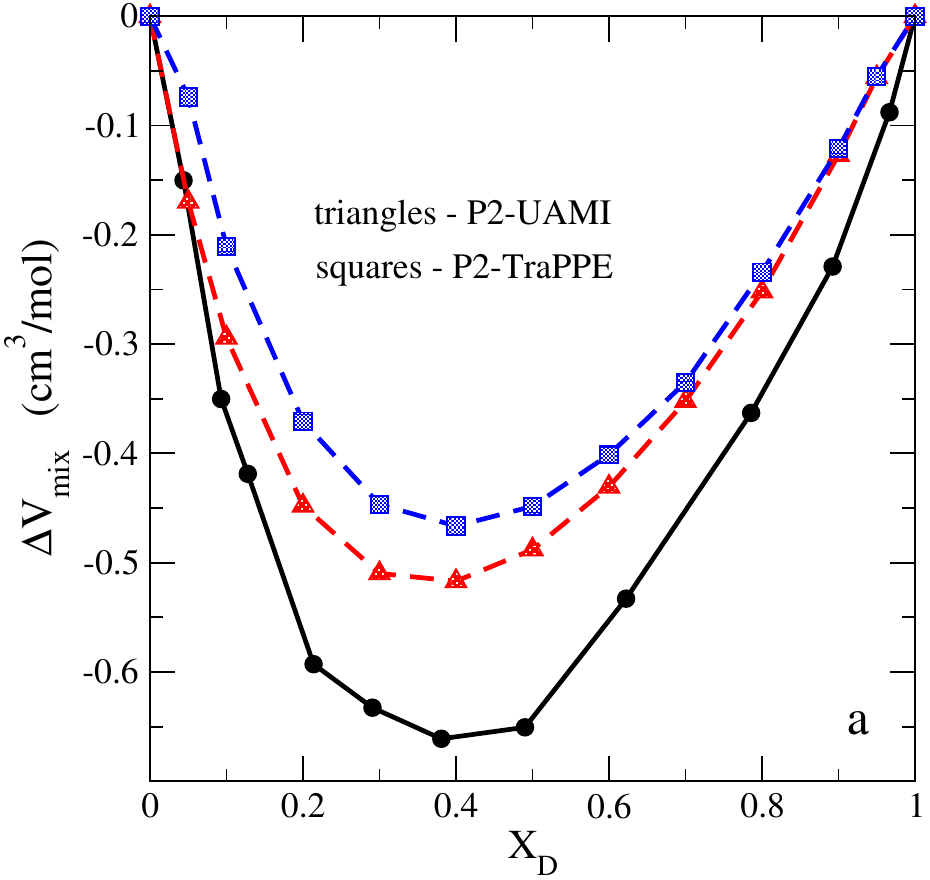}
\includegraphics[width=6cm,clip]{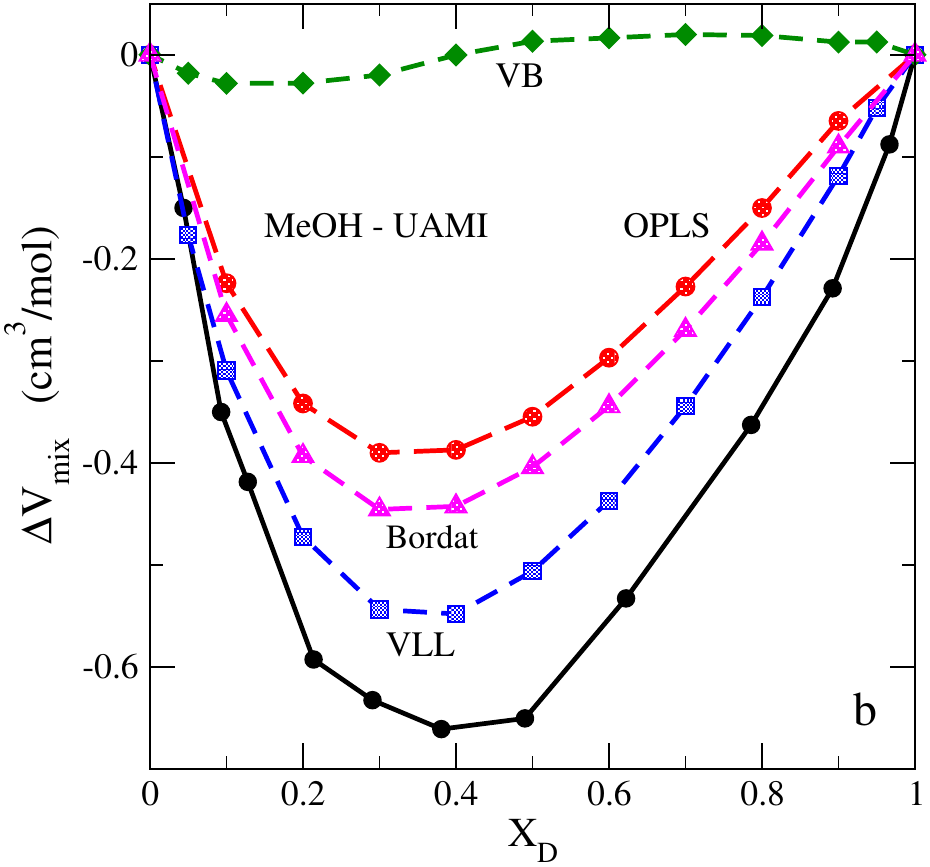}
\end{center}
\caption{(Colour online) Panels a and b:
The dependence of DMSO--MeOH excess molar volume
on composition. 
The simulation results are for the P2--UAMI and for P2--TraPPE united atom model
in panel a. In panel b, the simulation data are for
OPLS--UAMI, VLL--UAMI,  VB--UAMI and Bordat--UAMI united atom models.
The experimental data (black circles) are from~\cite{nikam} in both panels. The nomenclature of colors and symbols is
given in the figures.
}
\protect
\label{fig3}
\end{figure}

This representation
yields enhanced insight into the accuracy of $\rho(X_{\text D})$ curves.
Our simulation data for different models and comparison with experimental results
are shown in figure~\ref{fig3}. The curves in panel a of this figure provide insight of
the effect coming from changes of MeOH model whereas panel b illustrates
the effect of changes of $\Delta V_{\text{mix}}$ for different DMSO models.
From panel a of this figure we conclude that both MeOH models combined with 
P2 DMSO provide reasonable results for the excess mixing volume. The UAMI model
is slightly better than the TraPPE one for this specific property.
Panel b of figure~\ref{fig3} illustrates quality of the curves for $\Delta V_{\text{mix}}(X_{\text D})$
upon changes of the DMSO models for a given MeOH force field. These results
confirm that the VLL--UAMI model yields quite accurate predictions.
By contrast, the VB--UAMI model is not satisfactory for the excess mixing volume.
This behavior is due to the differences in the choice of parameters of Lennard-Jones
interaction as well as due to the details of intramolecular structure, 
because the charge distribution for VLL and VB models is the same.
In summary, one can use various united atom models for DMSO--MeOH mixtures to 
describe geometric aspects of mixing (in terms of excess density and mixing volume) 
accurately.  Additional test of different models
involves calculation of the excess mixing enthalpy. We refer to this property as to
the descriptor of the energetic aspects of mixing.

\subsection{Energetic aspects of mixing}

\begin{figure}[h]
\begin{center}
\includegraphics[width=6.7cm,clip]{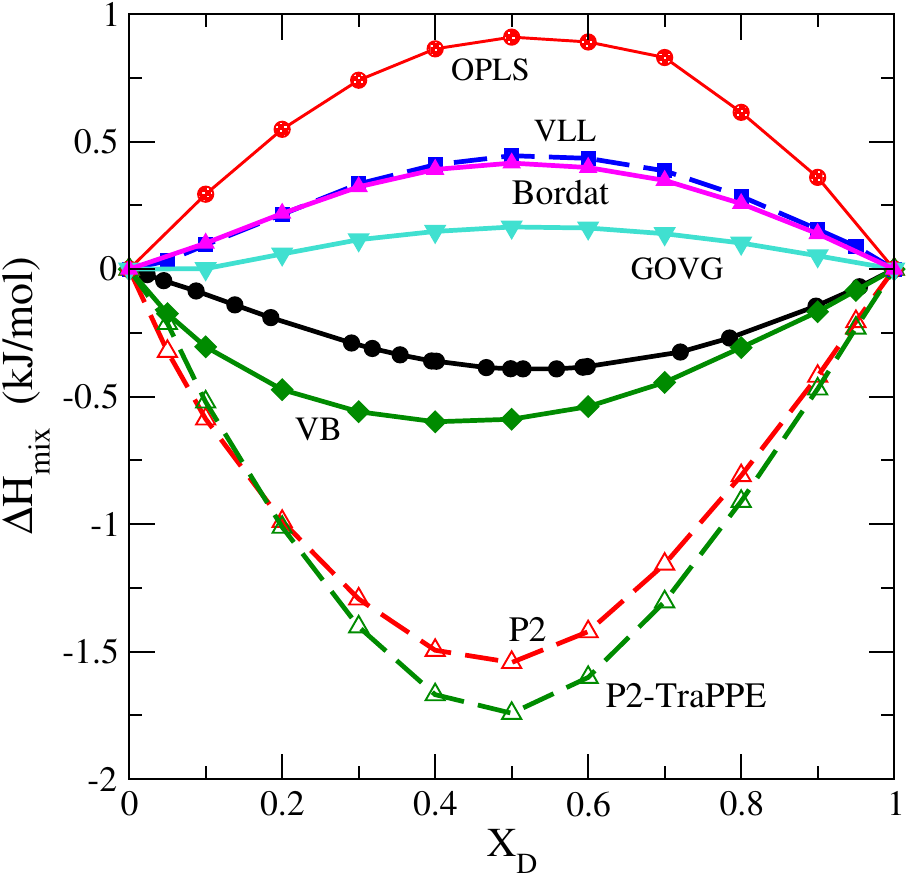}
\end{center}
\caption{(Colour online) 
The dependence of the excess mixing enthalpy on composition for
various models of DMSO--MeOH mixtures (indicated in the figure).
The experimental data (black circles) are from \cite{kimura}.
}
\protect
\label{fig4}
\end{figure}

In figure~\ref{fig4}, we present the  simulation results for excess mixing enthalpy and
make comparison with experimental data. It is a very sensitive property, to capture
it accurately presents a real challenge for the force field.

We observe that the models of the present study are not entirely satisfactory.
Some of them predict endothermic heat of mixing in contrast to exothermic mixing
from the experimental data. The best agreement with experimental results is
observed from the VB--UAMI model. The worst behavior, with correct sign but
in terms of absolute values, comes from P2--UAMI and P2--TraPPE models.
Smaller deviation from the experimental data,  but with incorrect sign, is obtained,
if the VLL, Bordat, GOVG DMSO models are combined with UAMI force field for MeOH.
This behavior is slightly disappointing. Actually, it should be attributed
to inaccurate description of the energetics of cross interaction between species
or more precisely to the description of energetic balance between interaction of
similar and dissimilar species.

In order to get a deeper insight into these trends, we have undertaken calculations of the
structural aspects of mixing in terms of pair distribution functions and
of evolution of the number of hydrogen bonds in the system upon changes of
composition.

\subsection{Structural aspects of mixing}

To begin with, we present the pair distribution functions of oxygens belonging
to methanol and DMSO at methanol-rich ($X_{\text D} = 0.05$), DMSO-rich ($X_{\text D} = 0.95$) 
and intermediate composition ($X_{\text D} = 0.5$) of the mixture in the following figures.
Two models, P2 and VLL, for DMSO have been taken because they yield a rather different
behavior of $\Delta H_{\text{mix}}(X_{\text D})$ in figure~\ref{fig4}. Thus, our principal issue is to elucidate the 
differences in the evolution of microscopic structure upon changes of DMSO content in
the model mixtures. 

\begin{figure}[h!]
\begin{center}
\includegraphics[width=5.5cm,clip]{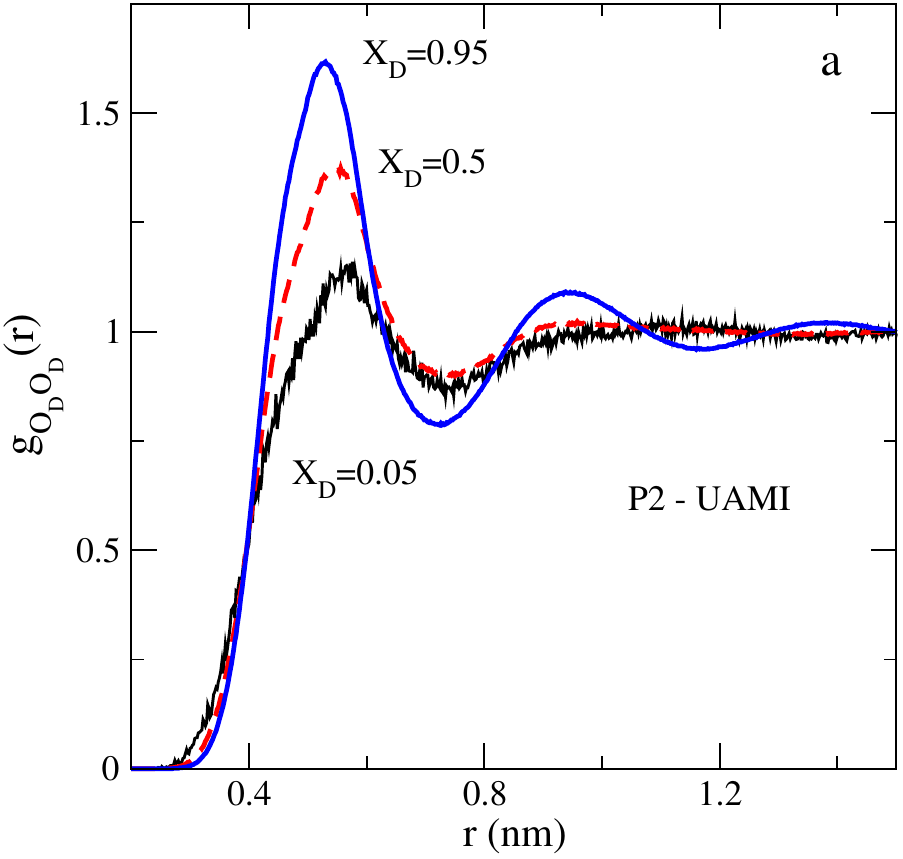}
\includegraphics[width=5.5cm,clip]{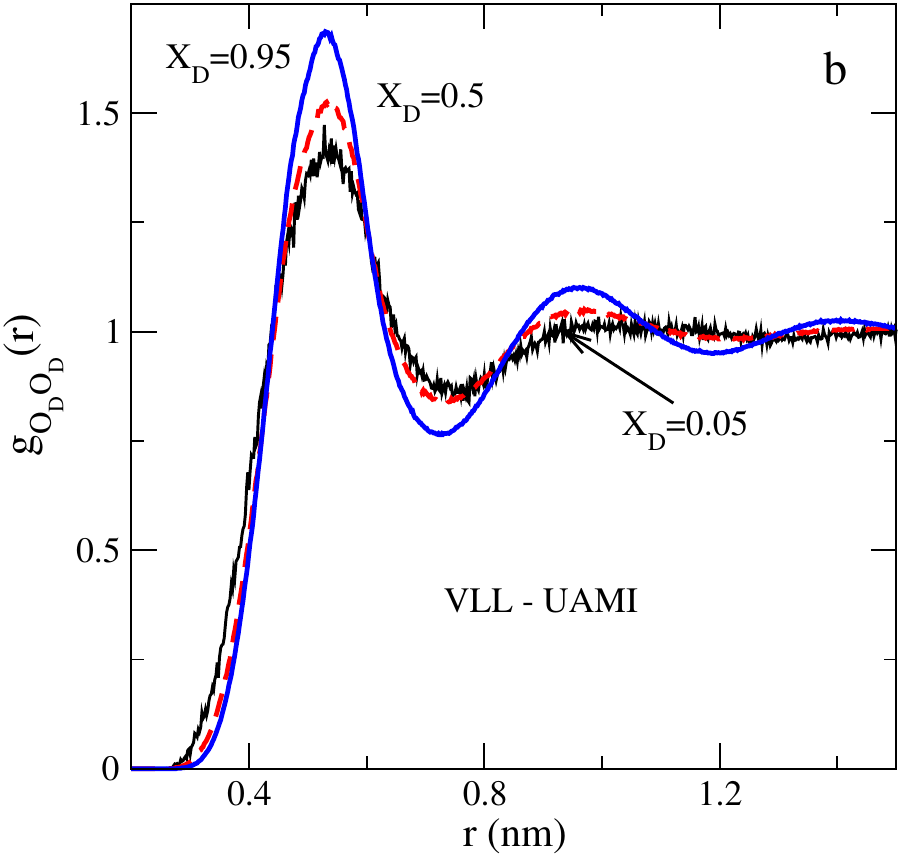}
\end{center}
\caption{(Colour online)
Evolution of the distribution of the DMSO oxygens, O$_{\text D}$O$_{\text D}$, on the mole fraction of
DMSO species in the mixture.
}
\protect
\label{fig5}
\end{figure}

Concerning the O$_{\text D}$-O$_{\text D}$ distribution, we observe that the development of the
short-range structure is qualitatively similar in the P2--UAMI and VLL--UAMI models, figure~\ref{fig5}.
Certain quantitative difference can be seen in the rate of growth of the first maximum
of the distribution function when $X_{\text D}$ increases from 0.05 to 0.5, however. Subtle
differences in the changes of structure might develop due to the differences
of parameters of Lennard-Jones direct interaction and dipole-dipole intermolecular potential,
besides correlation via methanol species. 

\begin{figure}[h!]
\begin{center}
\includegraphics[width=5.5cm,clip]{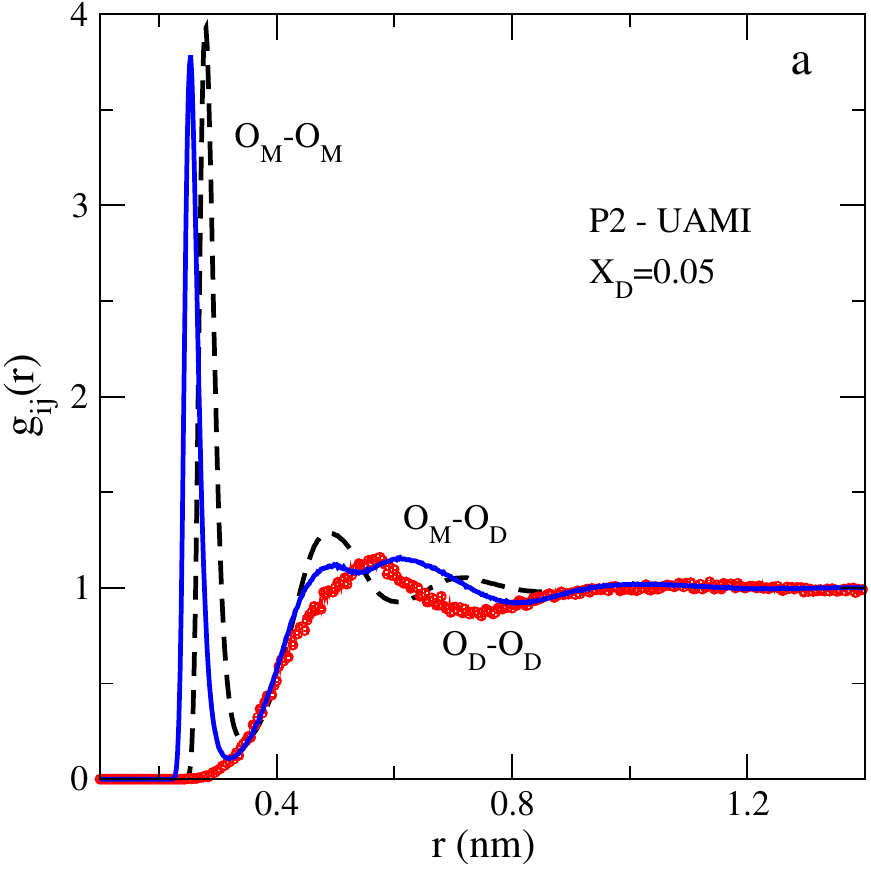}
\includegraphics[width=5.5cm,clip]{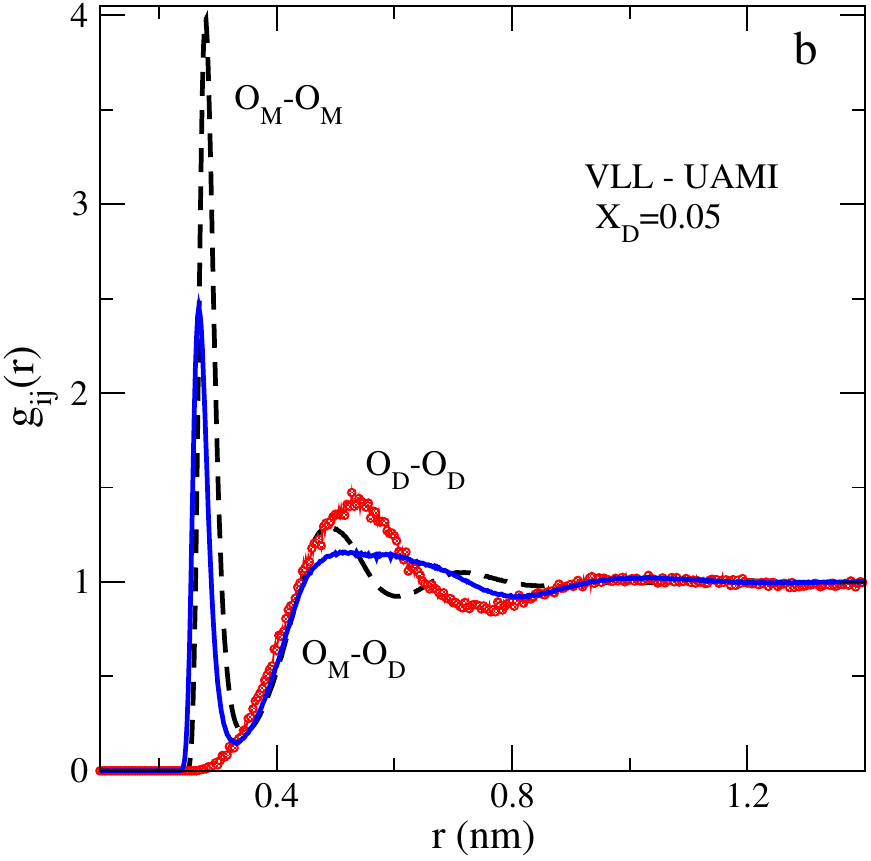}
\end{center}
\caption{(Colour online)
Pair distribution functions of oxygens belonging to methanol species, O$_\text{M}$O$_\text{M}$ (black dashed lines), 
cross, O$_\text{M}$O$_{\text D}$ (blue solid lines), and
of oxygens of DMSO molecules O$_{\text D}$O$_{\text D}$ (red lines) at $X_{\text D}= 0.05$. 
Panel a refers to P2--UAMI model whereas
panel b is for VLL--UAMI model. The same nomenclature of colors is used in
figures~\ref{fig6}--\ref{fig8}.
}
\protect
\label{fig6}
\end{figure}

\begin{figure}[h!]
\begin{center}
\includegraphics[width=5.5cm,clip]{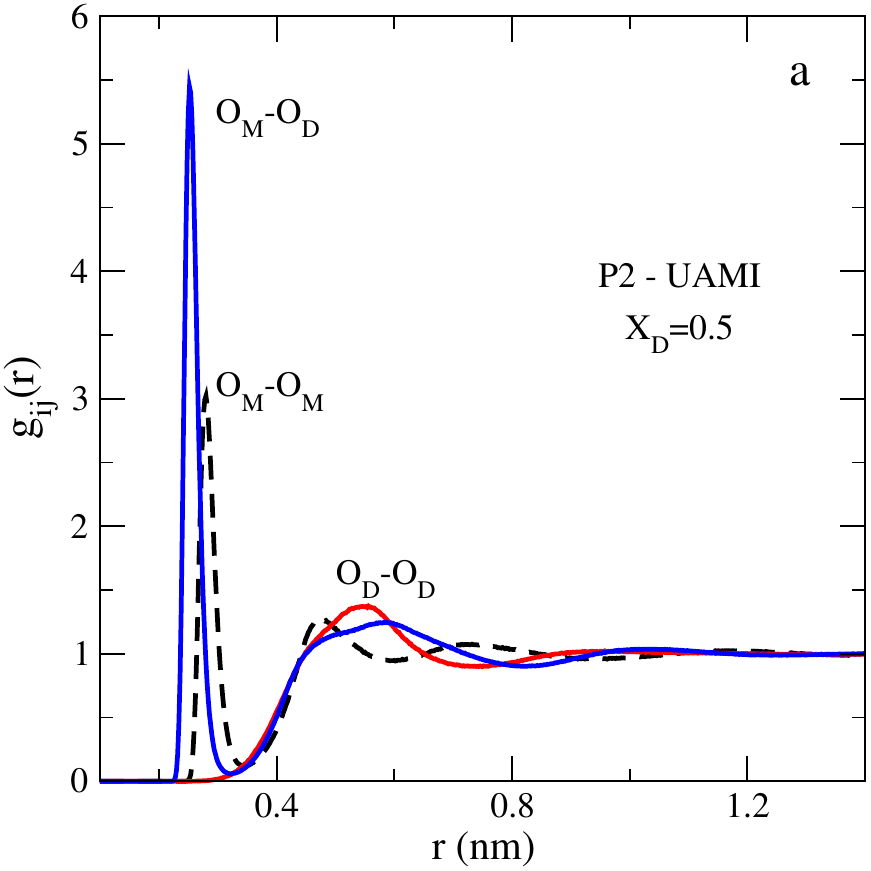}
\includegraphics[width=5.5cm,clip]{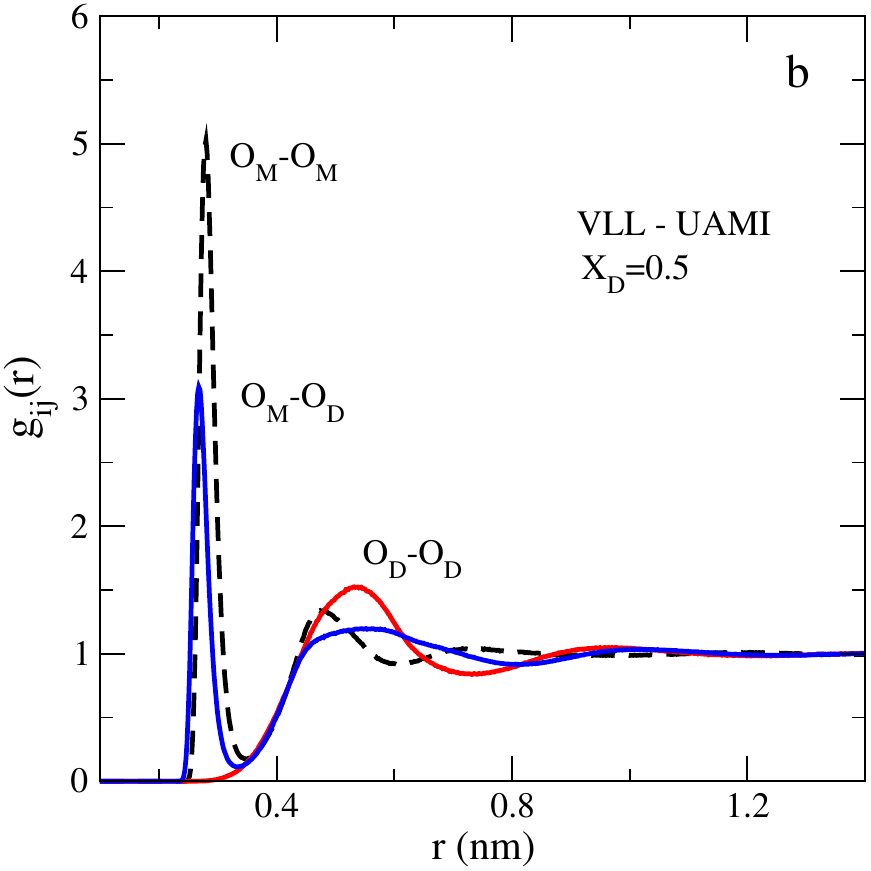}
\end{center}
\caption{(Colour online)
The same as in figure~\ref{fig6} for the mixture with equimolar composition ($X_{\text D}= 0.5$).
}
\protect
\label{fig7}
\end{figure}

\begin{figure}[h!]
\begin{center}
\includegraphics[width=5.5cm,clip]{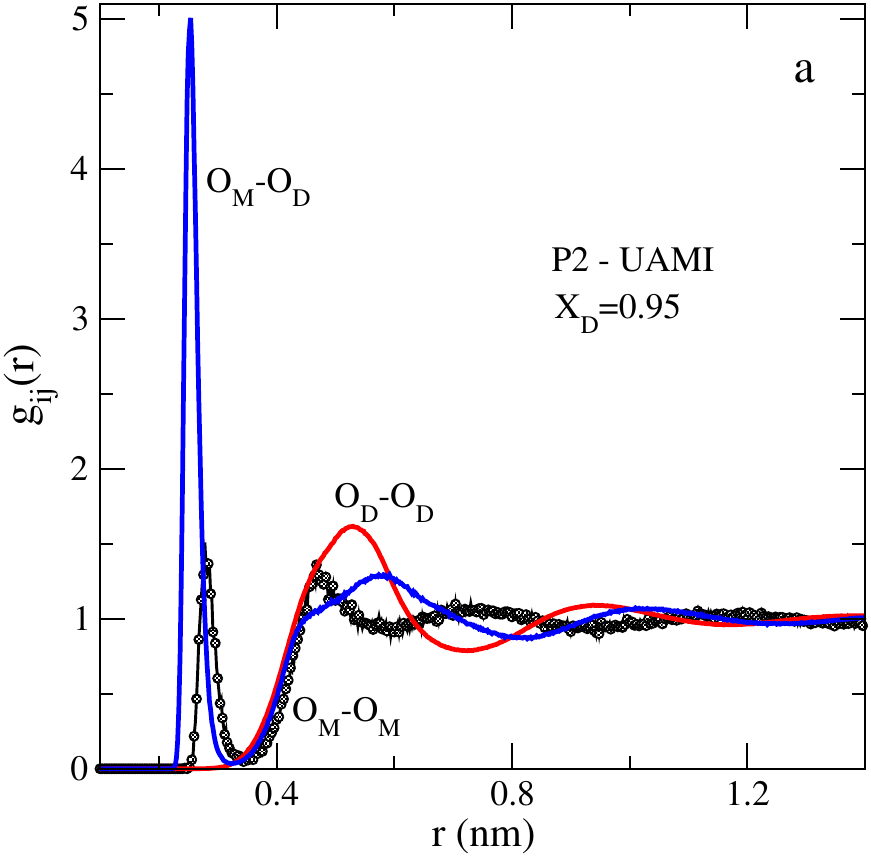}
\includegraphics[width=5.5cm,clip]{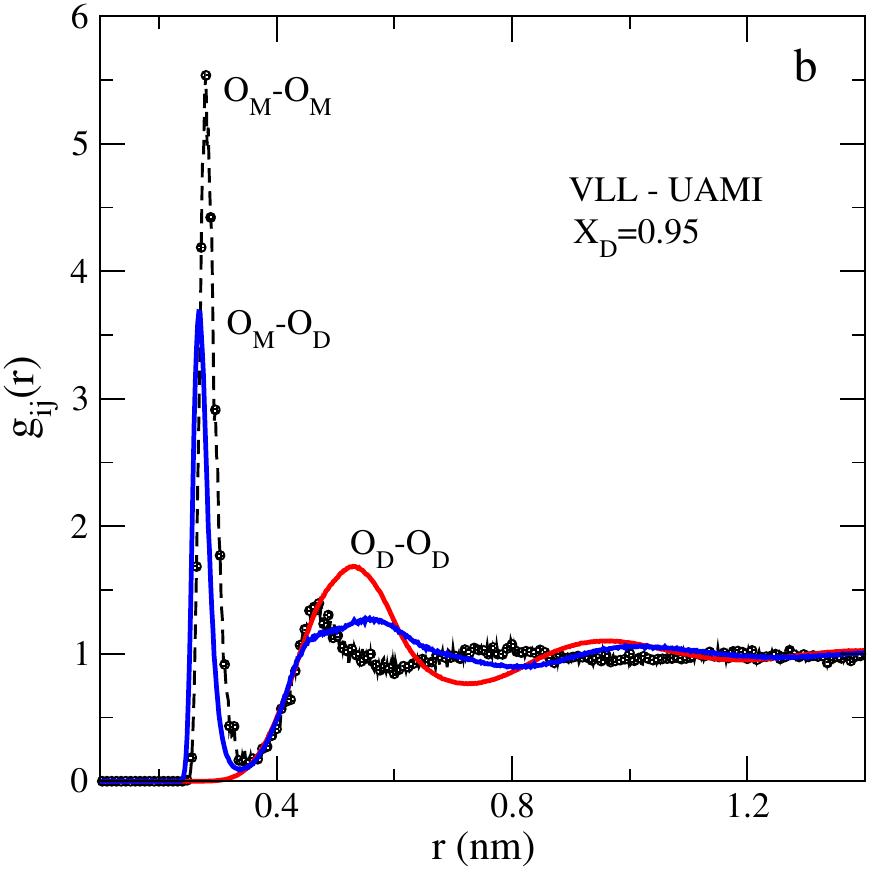}
\end{center}
\caption{(Colour online)
The same as in figure~\ref{fig6} and figure~\ref{fig7} but for DMSO-rich mixture with $X_{\text D}= 0.95$.
}
\protect
\label{fig8}
\end{figure}

Evolution of the cross pair distribution function between oxygens belonging to
different species, O$_\text{M}$O$_{\text D}$, exhibit similar trends for both models in question as well,
figures~\ref{fig6}--\ref{fig8}.
Specifically, the height of the first maximum of this function increases with 
increasing $X_{\text D}$ from 0.05 up to 0.5, figure~\ref{fig6} and \ref{fig7}. Next, it saturates and even decreases slightly
for the P2--UAMI models, whereas for the VLL--UAMI one, the first maximum keeps growing,
cf. figure~\ref{fig6} and figure~\ref{fig7}.

The principal difference in changes of the microscopic structure of two models can be
observed in terms of O$_\text{M}$O$_\text{M}$ distribution, figures~\ref{fig6}--\ref{fig8}. At a low DMSO concentration,
i.e., in methanol-rich mixtures, two models confirm very similar and strong correlation between 
MeOH molecules, figure~\ref{fig6}. Apparently, the DMSO molecules show stronger affinity to methanol
particles in the P2--UAMI model compared to the VLL--UAMI one, as it follows from the
first maximum of O$_\text{M}$O$_{\text D}$ distribution, figure~\ref{fig6}.

However, the height of the first maximum of O$_\text{M}$O$_\text{M}$ distribution gradually decreases upon increasing
$X_{\text D}$ within the P2--UAMI model, cf. figure~\ref{fig7}a and figure~\ref{fig8}a. By contrast, 
this maximum gradually grows with increasing $X_{\text D}$ within the VLL--UAMI model, cf. figure~\ref{fig7}b and
figure~\ref{fig8}b. This behavior is accompanied by increasing strength of affinity of O$_\text{M}$ and O$_{\text D}$
atoms, as it follows from changes of O$_\text{M}$O$_{\text D}$ distribution. Nevertheless,
the magnitude of correlations between dissimilar species, O$_\text{M}$ and O$_{\text D}$, is quite
different for two models, witnessed by the height of the first maximum of O$_\text{M}$O$_{\text D}$ distribution
in panels a and b of these three figures, figures \ref{fig6}--\ref{fig8}.

It is difficult to extract a more detailed information concerning changes of 
the microscopic structure and interpret predictions from the pair distribution functions
at united atom level modelling without additional assumptions. One needs to assume
the all-atom models to perform comparisons with the experimentally reachable
structure factors. We refer to~\cite{bako1,galicia1,mendez1} for detailed discussion of
this issue. Unfortunately, we are not aware of the experimental measurements
of the structure factors either from X-ray of neutron diffraction techniques. 

\begin{figure}[h!]
\begin{center}
\includegraphics[width=6.5cm,clip]{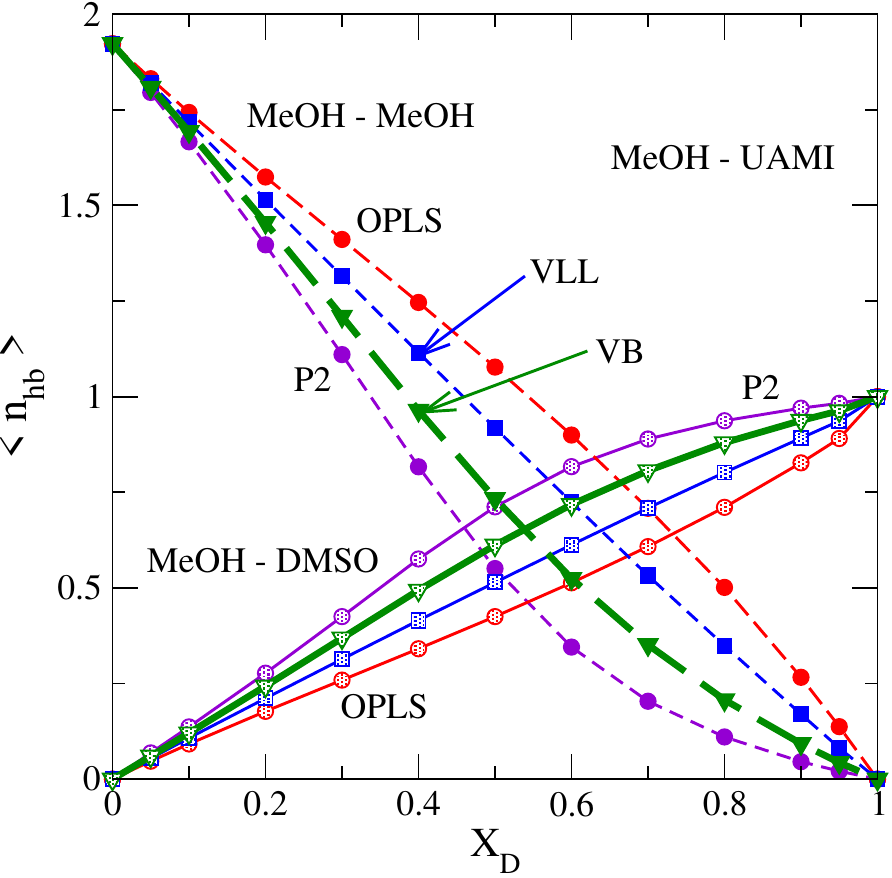}
\end{center}
\caption{(Colour online)
The dependence of the average number of hydrogen bonds on composition for
various models of DMSO--MeOH mixtures (indicated in the figure).
Solid lines describe cross hydrogen bonding,
i.e., between DMSO and MeOH molecules, whereas the dashed lines illustrate
bonding between methanol species.
}
\protect
\label{fig9}
\end{figure}

Additional insights into the structure of the mixtures in question follow
from the analysis of hydrogen bonding. The average number of
hydrogen bonds per methanol molecule  was obtained using the gmx hbond
utility with default options (it implies geometric criterion for H-bonding) 
of the GROMACS software. The reader can find a more detailed description
of the criteria and methods used to characterize hydrogen bonding effects 
in~\cite{skinner,zhang,galicia2}.
Our results are shown in figure~\ref{fig9}. 

General trends of behavior of the number of molecules participating in hydrogen bonds
on the DMSO mole fraction are quite similar from model to model. Namely, the 
average number of methanol molecules that form H-bonds between themselves, $\langle n_{hb}\rangle_\text{MM}$, decreases
with increasing the DMSO content in the mixture. On the other hand, the number of
cross hydrogen bonds between methanol and DMSO
molecules,  $\langle n_{hb}\rangle_\text{MD}$, increases with increasing $X_{\text D}$. At a low DMSO content,
the methanol molecules form two H-bonds between themselves, proving the formation of
a certain number of methanol chains in the system under such conditions.
On the other hand, in the DMSO-rich mixtures, the methanol molecules form one bond on average
with DMSO species while MeOH--MeOH bonds are improbable.

Interestingly, the balance between the numbers describing bonding between similar and dissimilar
species is different for different models. To keep track with the performance of the models
in question for the excess mixing enthalpy, we would like to comment the bonding in P2- UAMI
and VLL--UAMI models. 
In the former case, P2--UAMI, the $\langle n_{hb}\rangle_\text{MM}$ drops most rapidly with increasing 
$X_{\text D}$, compared to other models, e.g., the VLL--UAMI. On the other hand, the $\langle n_{hb}\rangle_\text{MD}$ 
increases most fast upon $X_{\text D}$ within P2--UAMI model, compared to
other models, figure~\ref{fig9}. However, at a high DMSO content, one can observe trends for 
$\langle n_{hb}\rangle_\text{MD}$ saturation, in contrast to other models under study. 
The most pronounced deviation (of opposite sign) of the excess mixing enthalpy 
for the models of our attention, P2 and VLL,
from experimental results occurs at intermediate composition, X$_{\text D} \approx 0.5$, cf. figure~\ref{fig4}.
At such composition, the average number of cross bonds, $\langle n_{hb}\rangle_\text{MD}$ , is larger than
the $\langle n_{hb}\rangle_\text{MM}$ for P2--UAMI model, figure~\ref{fig9}. By contrast, we observe that
$\langle n_{hb}\rangle_\text{MD}$ is substantially lower than $\langle n_{hb}\rangle_\text{MM}$ for VLL--UAMI model 
in this interval of intermediate composition. Indeed, the balance of bonding of species 
in the mixture is reflected in the model predictions of excess mixing enthalpy.
As a finishing touch of this subsection, we would like to attract attention of the
reader to the experimental findings concerned with the role of methyl groups in
hydrogen bonding \cite{yu1}. An all-atom modelling of methanol and DMSO is required to
address the issues resulting from these experiments. 
A comparison of spectra obtained via auto-correlation functions  
at a different level of modelling, would provide a more detailed insight
into hydrogen bonding for the mixture in question.

\subsection{Static dielectric constant}

\begin{figure}[h]
\begin{center}
\includegraphics[width=6cm,clip]{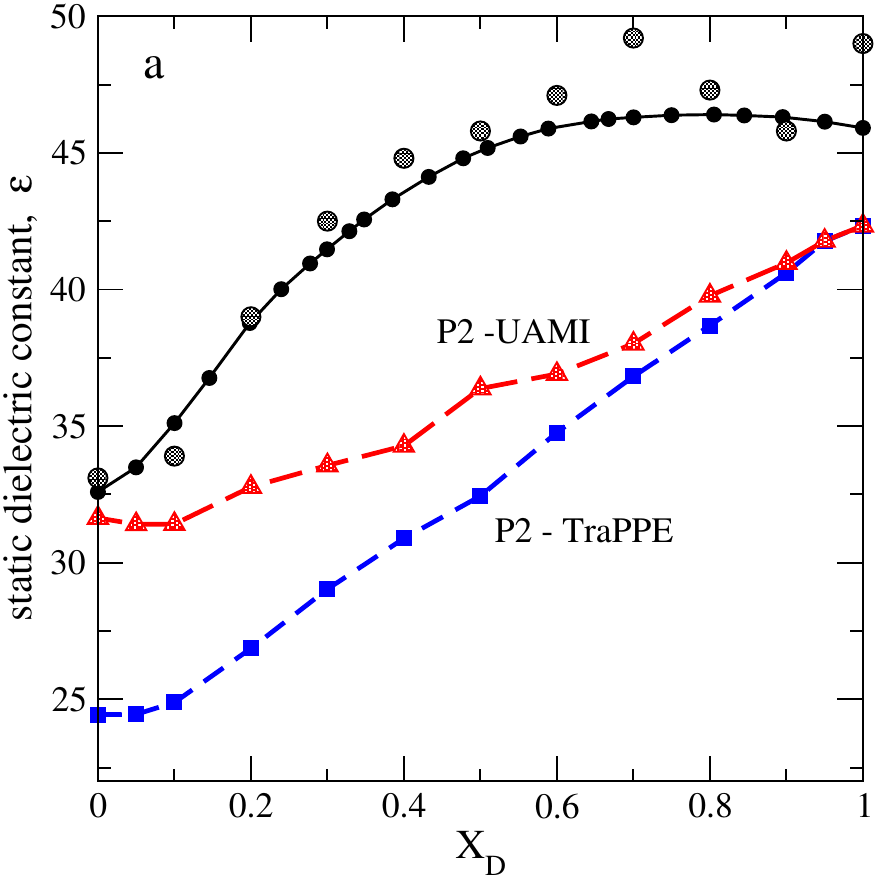}
\includegraphics[width=6cm,clip]{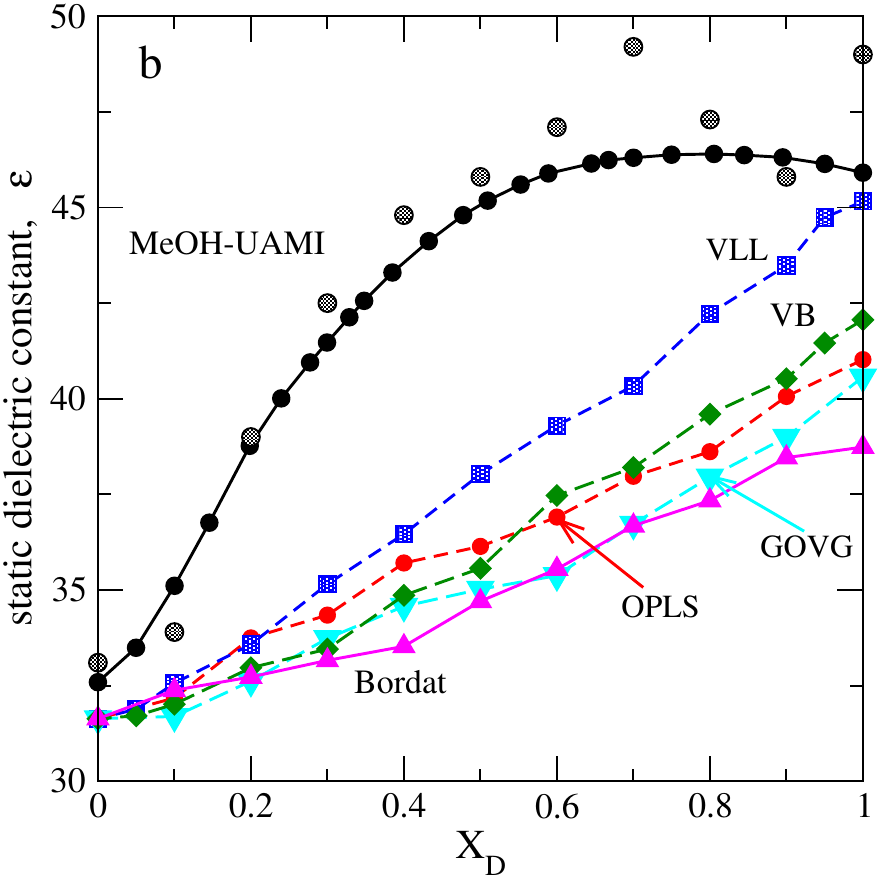}
\end{center}
\caption{(Colour online)
The dependence of the dielectric constant on composition for
various models of DMSO--MeOH mixtures (indicated in the figure).
The experimental data are from  \cite{roman} (solid black circles)
and \cite{guo} (dashed black circles).
}
\protect
\label{fig10}
\end{figure}

Our next concern is the description of the behavior of static dielectric constant, $\varepsilon$, 
for DMSO--MeOH mixture on composition. 
It is calculated by using common routine. Namely, $\varepsilon$ is obtained
from the time-average of the fluctuations of the total
dipole moment of the system~\cite{martin},

\begin{equation}
\varepsilon=1+\frac{4\piup}{3k_\text{B}TV}\big(\langle\bf M^2\rangle-\langle\bf M\rangle^2\big),
\end{equation}
where $k_\text{B}$ is the Boltzmann constant and V is the simulation cell volume.
The simulation results for two MeOH models combined with P2 DMSO, and different 
DMSO models combined with UAMI methanol are shown in figure~\ref{fig10}a and \ref{fig10}b, respectively.

We have already mentioned above that the principal improvement of the UAMI modelling
in comparison with TraPPE is in a better description of the dielectric constant 
of pure alcohols. This is illustrated in figure~\ref{fig10}a. The 
united atom P2 model for pure DMSO underestimates the dielectric constant, in comparison
with the experimental result. The P2--UAMI model  describes that the dielectric 
constant grows with increasing the DMSO content. 

In some sense, the curves constructed from the simulation results for 
different DMSO models behave similarly, figure~\ref{fig10}b.
The dielectric constant grows from the value for pure methanol (around 30) to a higher
value for the DMSO liquid. The VLL--UAMI leads to most reasonable values for DMSO-rich mixtures.
However, the shape of the curves for $\varepsilon(X_\text{D})$
resulting from experimental measurements and united atom type models is rather different.
The simulation data of all these models do not correctly reproduce the deviation from ideality
of $\varepsilon(X_\text{D})$. This behavior due to modelling should be explored
more in detail. Indeed, one needs to resort to all-atom models and to models that take into
account polarizability in order to get certain answers.  Particular attention should be paid to an 
appropriate description of cross hydrogen bonding between molecules belonging to different species. 
On pessimistic note, we are aware of the problem
known already for DMSO--water mixtures~\cite{gujt,aguilar,luzar-old}. 
It has not been solved so far even in the framework of one version of polarizable 
models \cite{bachmann}. 

\subsection{Self-diffusion coefficients of species}

\begin{figure}[h]
\begin{center}
\includegraphics[width=5.5cm,clip]{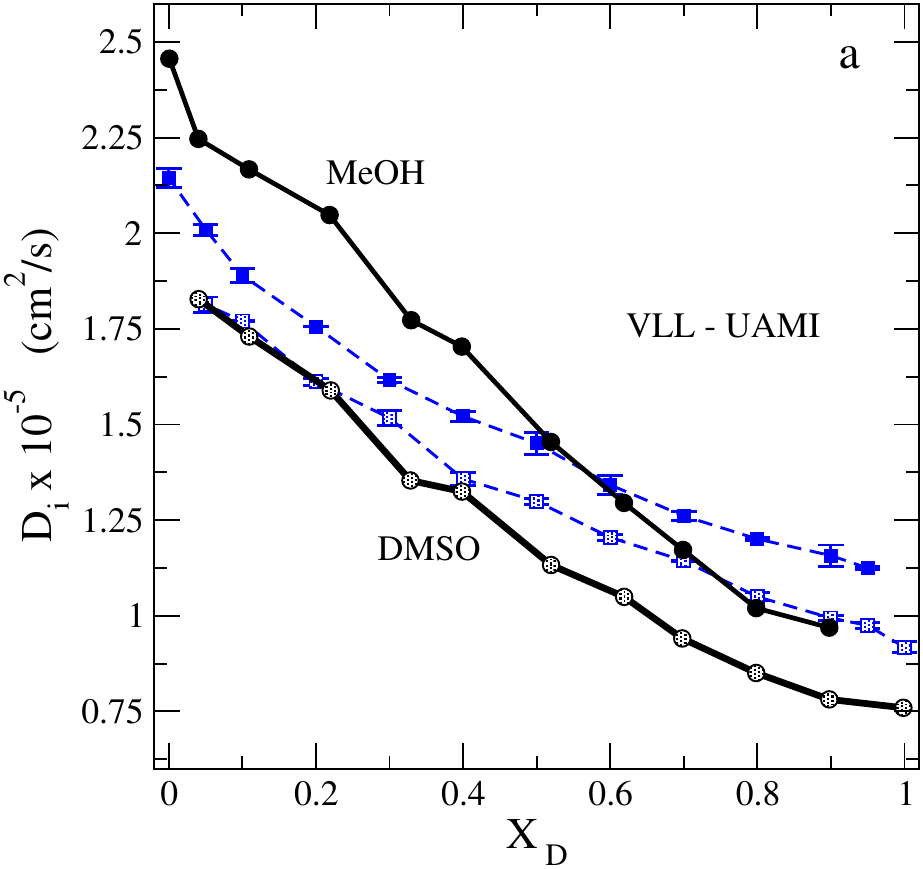}
\includegraphics[width=5.5cm,clip]{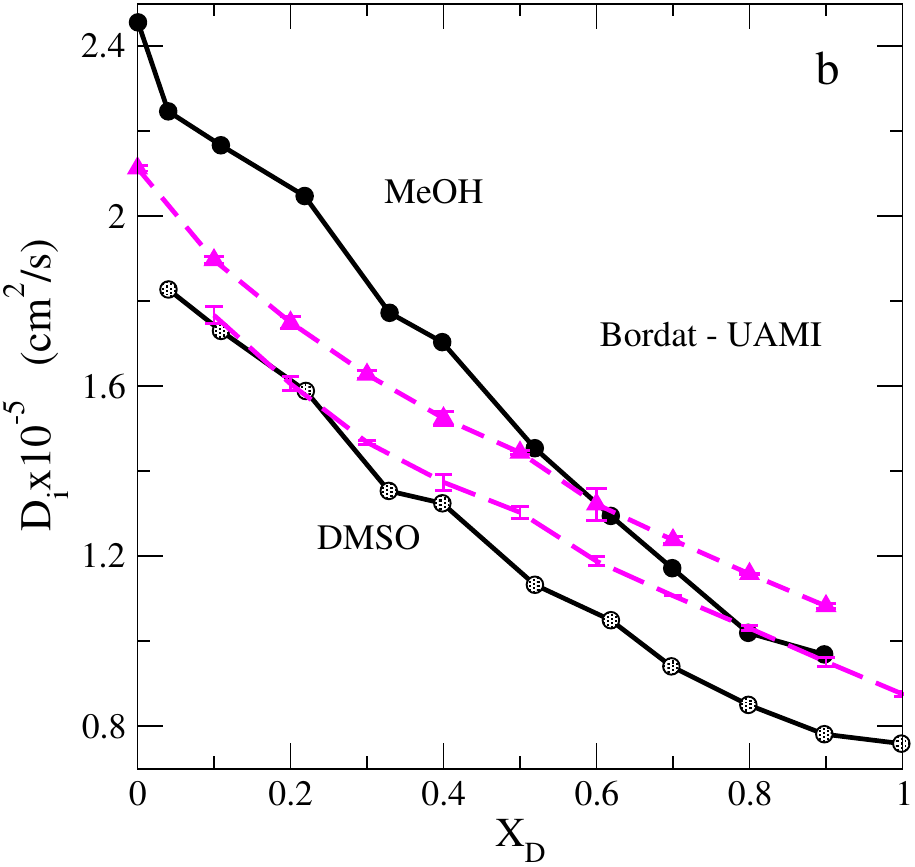}
\includegraphics[width=5.5cm,clip]{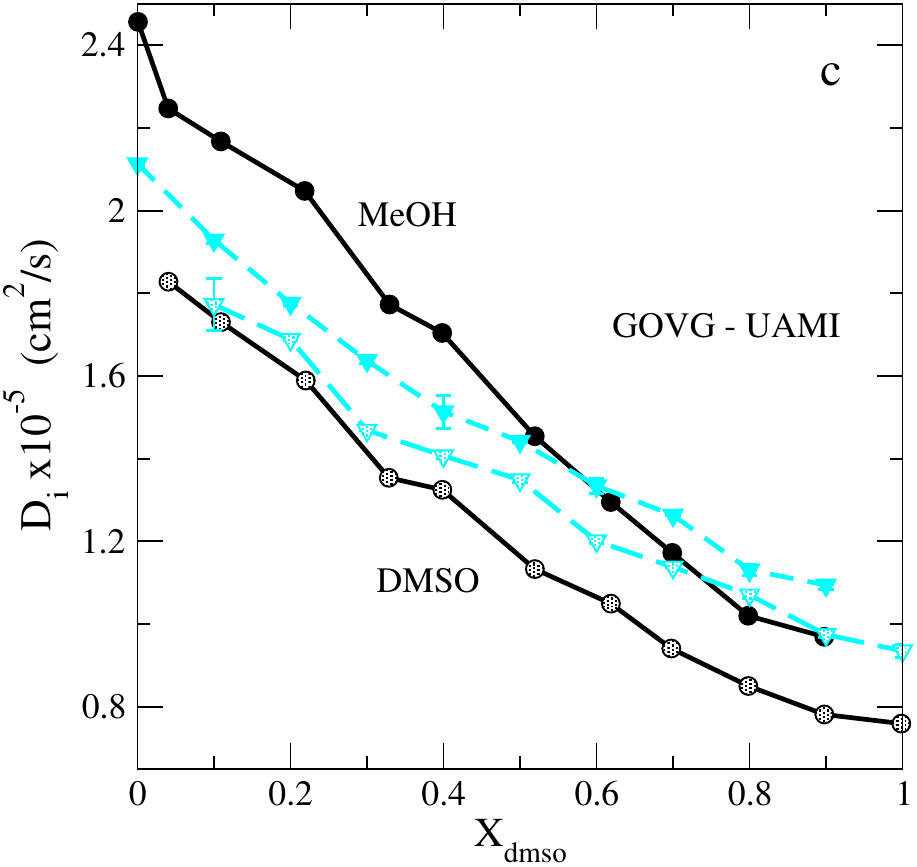}
\end{center}
\caption{(Colour online)
The dependence of the self-diffusion coefficients of species on composition for
DMSO--MeOH mixtures with VLL--UAMI, Bordat--UAMI and GOVG--UAMI models, respectively.
The nomenclature of colors for these models is like in figure~\ref{fig4}.
The experimental data (solid lines) are from \cite{virk}. The simulation data are
shown by dashed lines. Solid symbols are for MeOH whereas hollow symbols are for DMSO.
}
\protect
\label{fig11}
\end{figure}

Our results for the self-diffusion coefficients of methanol and DMSO species in the
mixtures of different composition are shown in figure~\ref{fig11}.
They follow straightforwardly from the Einstein relation,
\begin{equation}
D_i =\frac{1}{6} \lim_{t \rightarrow \infty} \frac{\rd}{\rd t} \vert {\bf r}_i(\tau+t)-{\bf r}_i(\tau)\vert ^2,
\end{equation}
where  $\tau$ denotes the time origin. Default settings of GROMACS were used for the separation of
the time origins and for the fitting interval. We performed a couple of tests generating center of mass
trajectories and used them to calculate the self-diffusion coeffients of species. However,
the obtained results were very similar to the outputs from the GROMACS utility for the system
under study.
On the other hand, the self-diffusion coefficients can be obtained from the auto-correlation
functions via Green-Kubo procedure. This method has not been used in the present work.

Unfortunately, the experimental data of this property are scarce. According to
the measurements given in \cite{virk}, the $D_{\text{MeOH}}$ is larger than
$D_{\text{DMSO}}$ in the entire composition range. The self-diffusion coefficients of
both species monotonously decrease upon increasing the DMSO content in the mixture.
Quality of the simulation results crucially
depends on the adequate description for pure components. 
The best performance
is provided by the VLL--UAMI, Bordat--UAMI and GOVG--UAMI models as seen in figure~\ref{fig11}.
Other two models suffer deficiencies yielding 
a crossover of the self-diffusion coefficients of
methanol and DMSO species, in contrast to experimental data (see figure~\ref{fig14}  in the Appendix).
We have not comprehensively explored the VB model for this property because 
it yields the self-diffusion coefficient for pure DMSO of the order of 2.5
(in close similarity to the model of Rao et al. as documented in 
table~\ref{tab3} of \cite{chalaris}).
In general, simulation results for the models illustrated in figure~\ref{fig11} are qualitatively
correct. In quantitative terms, the models underestimate the methanol self-diffusion
coefficient for methanol-rich mixtures and overestimate $D_i$ for both species,
methanol and DMSO, for DMSO-rich mixtures. There is enough room to improve the modelling
of this property considering it as one of principal targets rather than mere output.

\subsection{Shear viscosity of MeOH--DMSO mixtures}

Now we present some results concerned with the dependence of viscosity of 
model mixtures under study on composition.
We have chosen two models, VLL--UAMI and Bordat--UAMI,  that describe 
self-diffusion coefficients most appropriately according to figure~\ref{fig11}.
Previously, the shear viscosity from simulations for DMSO--MeOH mixtures was not discussed,
up to our best knowledge.  Experimental data for shear viscosity of
DMSO-alcohol mixtures are from \cite{nikam,baluja}.

The shear viscosity, $\eta$, is calculated according to the common routine 
described in various publications. Namely,
\begin{equation}
\eta = \frac{V}{kT} \int_0 ^\infty \rd t \left\langle  P_{\alpha \beta}(0) P_{\alpha \beta}(t) \right\rangle ,
\end{equation}
where, $\alpha, \beta$ = $xy, xz$ and $yz$. In order to evaluate the shear viscosity from the integral above,
we followed a procedure described in every detail by Gonzalez and Abascal, \cite{gonza}.
The statistical averages, given by points, follow from 3 - 6 independent simulations of
the components of pressure tensor.

\begin{figure}[h!]
\begin{center}
\includegraphics[width=5.5cm,clip]{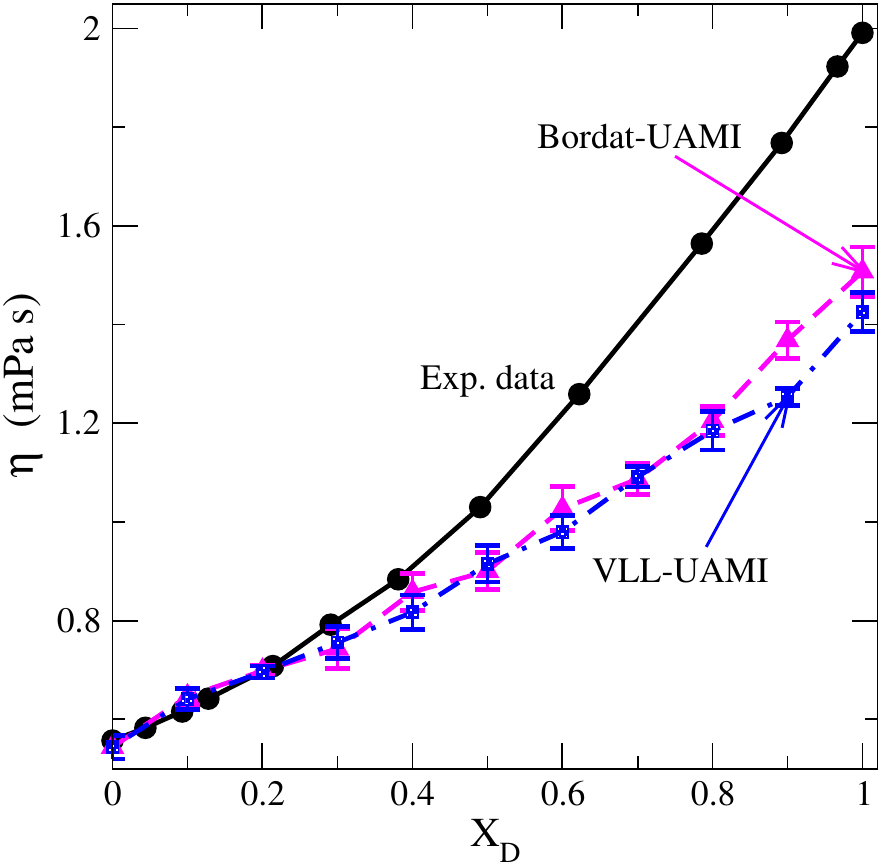}
\end{center}
\caption{(Colour online)
The dependence of shear viscosity on composition for
DMSO--MeOH mixtures with VLL--UAMI (blue squares) and
Bordat--UAMI (magenta triangles) models, respectively.
The experimental data (black solid circles) are from \cite{nikam}.
}
\protect
\label{fig12}
\end{figure}

Our simulation results are shown in figure~\ref{fig12}. The line that joins the calculated points is provided
for the sake of better visualization. The models in question describe shear viscosity 
reasonably well starting from methanol-rich mixtures up to intermediate composition in terms
of $X_{\text D}$. For DMSO-rich mixtures, the models under study underestimate the values for shear 
viscosity in comparison with experimental data. This feature of DMSO models is known 
already from the studies of water-DMSO systems, see, e.g., \cite{bagchi}.
The inaccuracy of the description of self-diffusion coefficients of species
for methanol-rich mixtures does not influence the accuracy of results for shear
viscosity. By contrast, overestimation of the self-diffusion coefficients 
of both species (M and D) for DMSO-rich mixtures, cf. figure~\ref{fig11}, results in the underestimation
of shear viscosity in this interval of composition.

\subsection{Surface tension of DMSO-MeOH mixtures on composition}

Our final remarks concern the behavior of the surface tension of DMSO-MeOH
mixtures. 
The  surface tension calculations at each
composition have been performed by taking the final configuration of particles
from the isobaric run. Next, the box edge along $z$-axis was extended by a factor of 3,
generating a rectangular box with liquid slab
and two liquid-mixture-vacuum interfaces in the $x-y$ plane,
in close similarity to the procedure applied in~\cite{vanderspoel}.
The total number of molecules is sufficient to yield an area of the
$x-y$ face of the liquid slab sufficiently big. The elongation of the liquid slab along
$z$-axis is satisfactory as well.
The effect of cutoff distance in the account of inter-particle interactions
has been discussed in detail in \cite{jose-cutoff}. Taking into account this analysis, 
in our calculations we used the cutoff distance equal to $2.2$ nm, 
i.e., twice larger than for the bulk simulations.

\begin{figure}[h!]
\begin{center}
\includegraphics[width=5.5cm,clip]{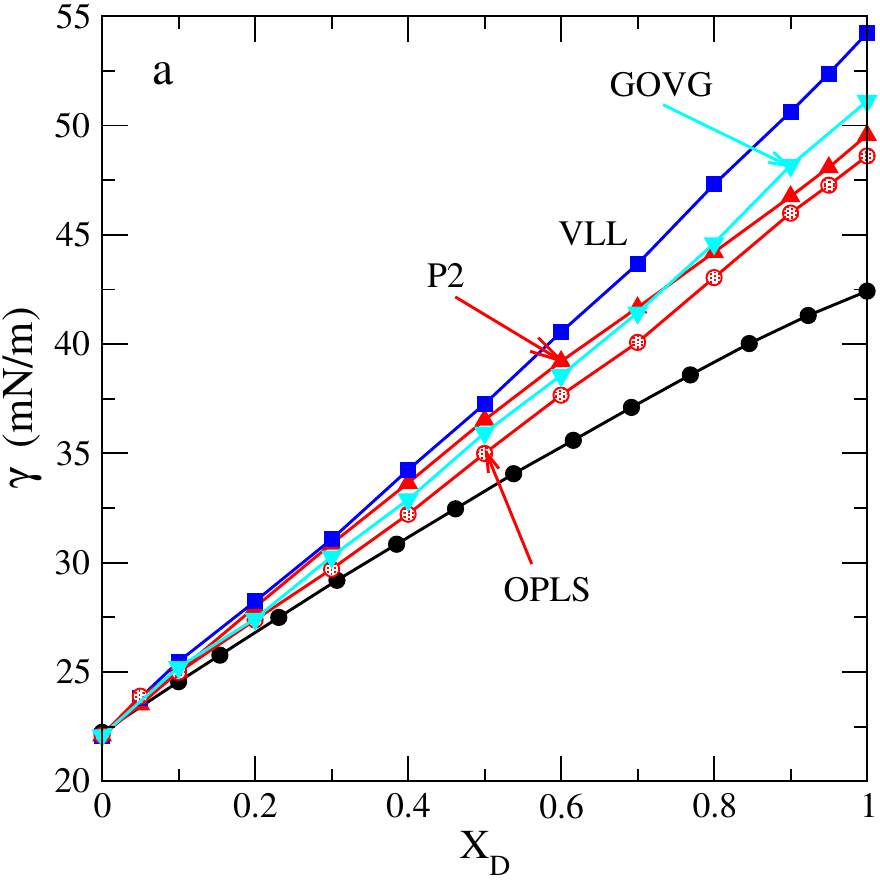}
\includegraphics[width=5.5cm,clip]{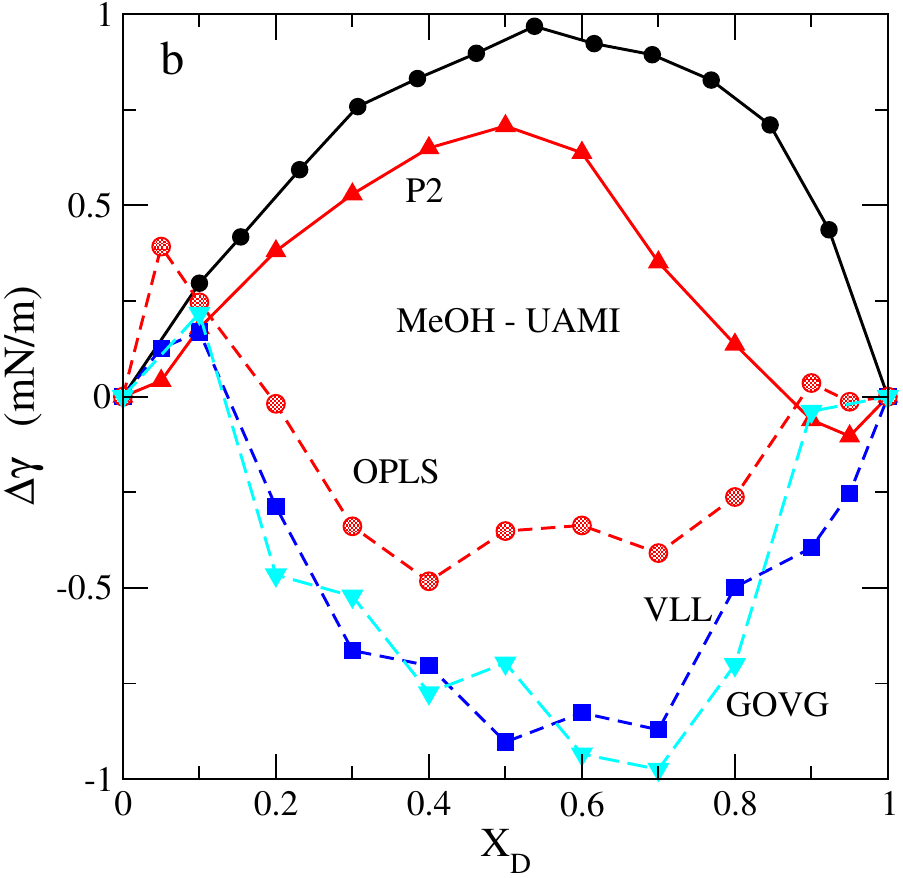}
\end{center}
\caption{(Colour online)
The dependence of the surface tension on composition (panel a)
and of the excess surface tension (panel b) for
DMSO--MeOH mixtures with UAMI methanol and various DMSO  models.
The experimental data are from \cite{bagheri}.
}
\protect
\label{fig13}
\end{figure}

The executable file has been modified by deleting the
fixed pressure condition while preserving the V-rescale thermostatting with the same parameters
as in the NPT runs. Moreover, the corrections for energy and pressure has not been
applied at this step of procedure. 
The values for the surface tension, $\gamma$, follow from the combination of the time
averages for the components of the pressure tensor,
\begin{equation}
\gamma = \frac {1}{2} L_z \Big\langle [P_{zz}-\frac{1}{2}(P_{xx}+P_{yy})]\Big\rangle,
\end{equation}
where $P_{ij}$ are the components of the pressure tensor along $i,j$ axes, and  $\langle\ldots\rangle$
denotes the time average.
We performed 9--10 consecutive runs at a constant volume, each piece 
with duration of 6~ns, and
obtained the result for $\gamma$ by averaging.
The experimental results were taken from \cite{bagheri}.

Principal feature of the simulation results is that all united atom models in question 
substantially overestimate the values for the surface tension of mixtures with dominating amount
of DMSO species. Agreement with experimental data is qualitative rather than quantitative. 
For methanol-rich mixtures, the quality of results is much better.
The reason of this behavior is that the alcohol model was parameterized with
the surface tension as a target, apart from density and dielectric constant \cite{melgarejo}.
By contrast, the DMSO united atom models from literature were parameterized
using the density and vaporization enthalpy as targets solely.
The curves for $\gamma(X_{\text D})$ from simulations seem to be almost linear. Therefore, 
the excess surface tension, $\Delta \gamma$,
values are small. Insufficient accuracy of the DMSO models in this respect
precludes capturing the deviation from ideality with reasonable precision.
The best description of $\Delta \gamma (X_{\text D})$ is provided by the
P2--UAMI model.

\section{Summary and conclusions}

Principal objective of this computer simulation work was to explore mixing properties of 
a mixture composed of a polar aprotic solvent (DMSO) and alcohol species
(amphiphilic molecules possessing both hydrophobic and hydrophilic groups).
We restricted our attention to methanol as the simplest alcohol.
Our study applies united atom models of these two species. Two models of methanol,
TraPPE and UAMI, are used. On the other hand, several models for DMSO are considered.
Namely, we use the  P2, OPLS, VLL, VB, GOVG and Bordat et al. models, see table~\ref{tab1}
in the models section.
Motivation of this ``low level'' modelling is that the available knowledge
of the properties of these systems from computer simulations is not comprehensive.
Moreover, theoretical predictions have not been confronted and tested 
with respect to a quite wide available experimental data for various properties.
The recently developed successful strategy for modelling the pure alcohols 
involved their density, the dielectric constant and surface tension as targets.
In general terms, this approach permitted to describe the self-diffusion coefficient,
shear viscosity, critical properties of pure alcohols and their solubility 
in water \cite{salas}. Applicability of this procedure to other systems
and a wider set of properties has not been fully proven so far.

On the other hand, the present day parametrization of the DMSO united atom models
involves the liquid density and vaporization enthalpy as targets, solely. In spite
of much efforts, using modification of the parameters of inter-particle interactions and
intra-molecular structure, the quality of results for several properties is not
perfect. In order to attempt a better modelling, one should have clear
vision of the problems necessary to solve for the description of DMSO-alcohol 
mixtures. We are aware of a more sophisticated modelling of DMSO (flexible all-atom,
polarizable, ab initio) proposed in, e.g.,~\cite{strader,bachmann,mancini}. However, 
much future work is necessary to systematically advance with this level of theory
to applications.

In this work, we explored a wide set of properties for DMSO--MeOH mixtures. They
include density and excess mixing density, molar volume and excess mixing volume,
excess mixing enthalpy, average number of hydrogen bonds of similar and dissimilar species,
dielectric constant, self-diffusion coefficients and shear viscosity, surface 
tension and excess surface tension. In all cases, we intended to critically
evaluate the results versus experimental data. A few aspects of the microscopic structure
are discussed as well. Unfortunately, the experimental counterpart, concerning the structural
properties on composition, is missing.
As concerns thermodynamic properties, such as self-diffusion coefficients, static dielectric
constant, surface tension and shear viscosity, for
practical use  we would recommend the
VLL, Bordat et al., GOVG models combined with UAMI methanol model. Indeed, they provide a quite reasonable agreement with experimental data
for various properties. However, the accuracy of some excess properties from simulations
of united atom models is not perfect.
Nevertheless, application of a set of these DMSO models with UAMI methanol
is recommended with necessary extensions, rather than to confide in the results from
a single, distinguished model.
Perhaps the findings of present work can be used to extend the training sets
and incorporate them into the protocols of machine learning methods for
studies of solubility of various compounds in DMSO as discussed in  \cite{tetko}.

Apparently, there are two principal problems to be solved. One of them is in a
better parametrization of pure DMSO considering different small sets of properties
as targets, in the spirit of parametrization procedure proposed for alcohols and more complex
molecules \cite{salas,nunez}. The second one lies in elucidation and appropriate
description of the cross interaction effects, i.e., of correlations between dissimilar species.
It seems that inclusion of the structural aspects of mixing into targets would be beneficial.
To implement this issue would require experimental studies via various diffraction techniques.
Unfortunately, we are not aware of the activities along this line of research.

Having in mind that the entire set of properties for the systems under study
follows from a balance of dipole-dipole interactions and association effects
leading to hydrogen bonding, it would be profitable to perform additional
calculations of the properties permitting interpretation of
spectroscopic results, see, e.g.,~\cite{bone}.  All these issues are under exploration in our
laboratory at present.

\section*{Acknowledgements}
 O.P. acknowledges helpful discussions with Dr. Laszlo Pusztai concerning various aspects of experimental research 
and interpretation of data for systems of the present study. 
M.C.S. is grateful 
to Dr. Carlos Vega for illuminating discussions concerned with calculations of shear viscosity.
We are grateful to Dr. Taras Patsahan for his interest to this work and his
important comments concerning the improvement of the manuscript.
We acknowledge a valuable technical support of this work by Magdalena Aguilar 
at Instituto de Quimica de la UNAM.

\appendix
\counterwithin{figure}{section}
\counterwithin{table}{section}

\section*{Appendix}
\setcounter{figure}{0}
\setcounter{section}{1}

Additional information for the composition dependence of self-diffusion coefficients.
	\setcounter{figure}{0}
\begin{figure}[h!]
\begin{center}
\includegraphics[width=5.5cm,clip]{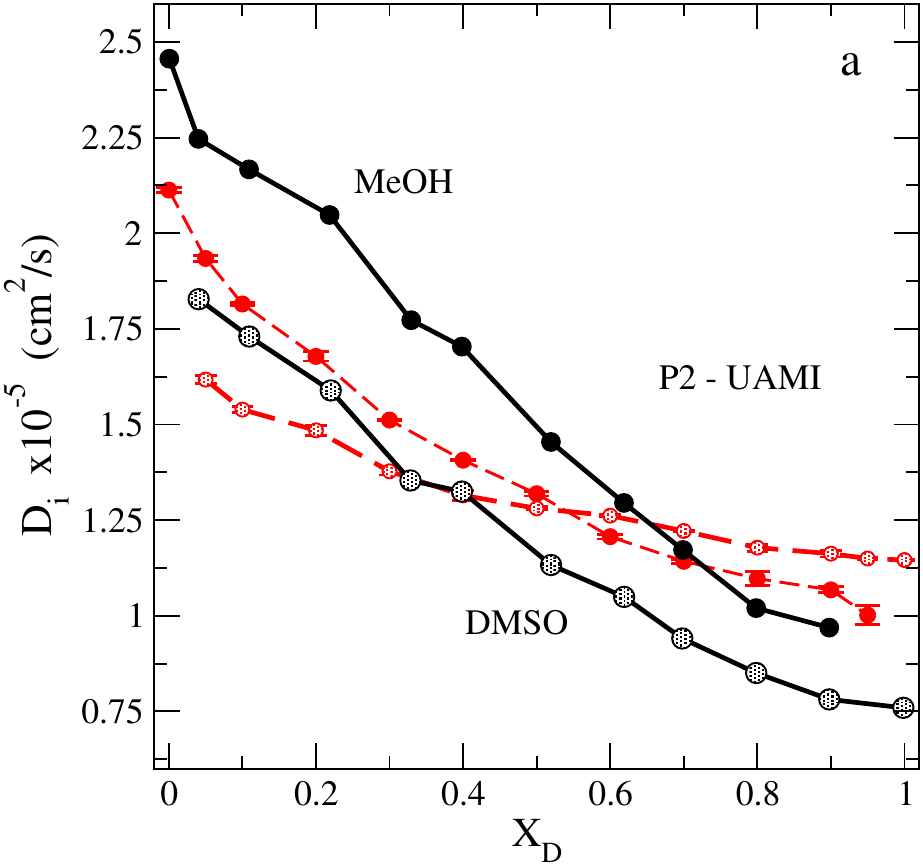}
\includegraphics[width=5.5cm,clip]{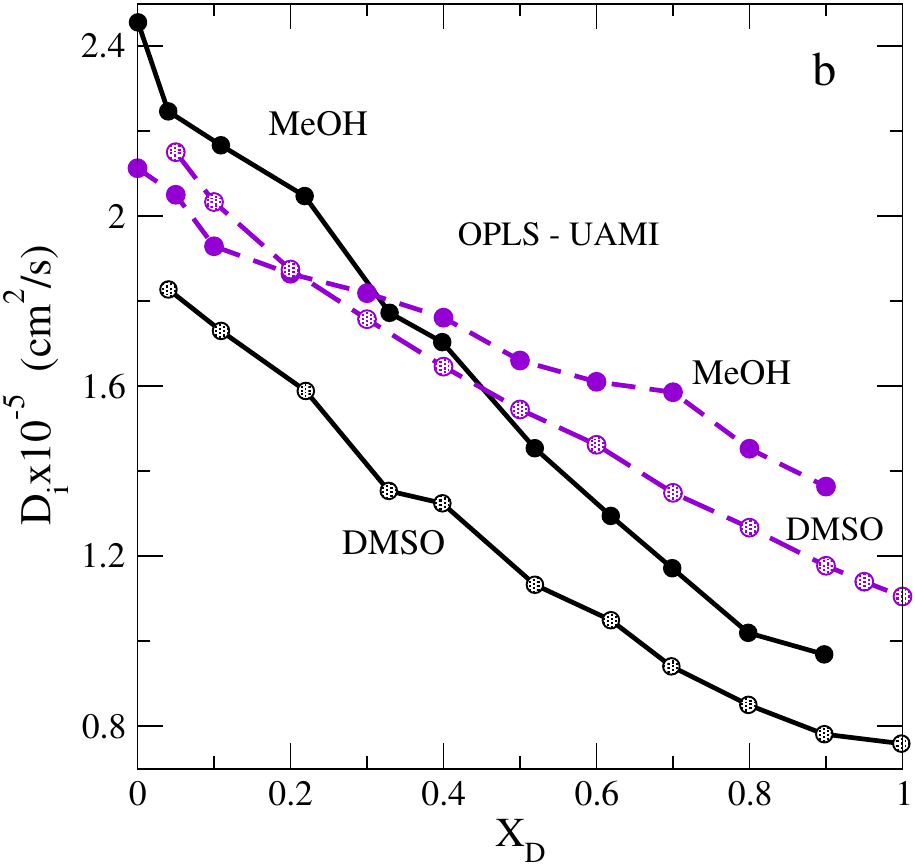}
\end{center}
\caption{(Colour online)
The dependence of the self-diffusion coefficients of species on composition for
DMSO--MeOH mixtures with P2--UAMI (panel a) and OPLS--UAMI (panel b) models.
The experimental data are from \cite{virk}. 
}
\protect
\label{fig14}
\end{figure}

Within these two models, one observes a crossover of the self-diffusion coefficients of
MeOH and DMSO species, in contrast to experimental data see figure~\ref{fig14}.

\ukrainianpart
\title[ДМСО -- одноатомні спирти]
{Моделювання рідких сумішей ДМСО--МеОН методом молекулярної динаміки.
	Вплив силових полів на властивості змішування 
}

\author
{М. Крус-Санчес\orcid{0009-0005-0407-6496}\refaddr{label1},
	O. Пізіо\orcid{0000-0001-8333-4652}\refaddr{label2}
	}
\addresses{
	\addr{label1}
	Хiмiчний факультет Автономного унiверситету Метрополiтана-Iстапалапа, просп. Сан Рафаель Атлiкско 186,  кол. Віцентіна, 09340, CDMX, Мехiко 
	\addr{label2}
	Інститут хiмiї, Нацiональний Автономний унiверситет Мехiко,
	Сіркіто Екстеріор, 04510, Мексика
}

\makeukrtitle
\begin{abstract}
	За допомогою комп'ютерного моделювання молекулярної динаміки вивчається залежність основних властивостей рідких сумішей диметилсульфоксиду (ДМСО)-метанолу (MeOH) від їх складу. 
	Досліджено набір неполяризовуваних напівгнучких моделей для молекули ДМСО в поєднанні з моделями метанолу. Оцінено тенденції зміни складу густини, надлишкового об'єму змішування та надлишкової ентальпії змішування.	Окрім того, вивчається залежність самодифузії частинок, зсувної в'язкості, статичної діелектричної проникності та поверхневого натягу від складу суміші. Проаналізовано певні аспекти мікроскопічної структури з точки зору радіальних функцій розподілу та середньої кількості молекул з водневими зв'язками.
	Якість багатьох комбінацій моделей проілюстровано та критично оцінено шляхом порівняння з експериментальними даними.
	\keywords молекулярна динаміка, метанол, диметилсульфоксид,
	густина, діелектрична проникність, поверхневий натяг, коефіцієнти самодифузії, зсувна в'язкість
\end{abstract}
\end{document}